\documentclass[lettersize,journal]{IEEEtran}

\IEEEoverridecommandlockouts
\usepackage{multirow}
\usepackage{bbding}
\usepackage{subfig}
\usepackage{bm}
\usepackage{stfloats}
\usepackage{makecell} 
\usepackage{dsfont}
\usepackage{threeparttable}
\usepackage{amsmath}
\usepackage{amsfonts}
\usepackage{color}
\usepackage{comment}
\usepackage{graphicx}
\usepackage{multirow}

\usepackage{url}
\usepackage{booktabs}  
\usepackage{enumitem}
\usepackage{float}
\usepackage{array}
\usepackage{float}

\usepackage[switch]{lineno}

\usepackage{lscape}
\usepackage{booktabs}
\usepackage{multicol}
\usepackage[ruled, noend]{algorithm2e}
\usepackage{xcolor}
\usepackage{colortbl}
\usepackage{wrapfig}
\usepackage{amssymb} 
\usepackage{cite}
\usepackage[breaklinks,colorlinks,citecolor=black]{hyperref}
\usepackage{CJK}
\definecolor{r1}{rgb}{0,0,0}
\definecolor{r2}{rgb}{0,0,0}
\definecolor{r3}{rgb}{0,0,0}
\definecolor{r4}{rgb}{0,0,0}

\newtheorem{Remark}{Remark}
\begin{document}
\begin{CJK}{UTF8}{gbsn}
\title{STABLE:$\,$Efficient$\,$Hybrid$\,$Nearest$\,$Neighbor$\,$Search$\,$via Magnitude-Uniformity and Cardinality-Robustness} 

\author{
        Qianyun Yang,~Zhiwei Chen,~Yupeng Hu~\IEEEmembership{Member,~IEEE}\thanks{Corresponding Author: Yupeng Hu},~Zixu Li,~Zhiheng Fu,~Liqiang Nie~\IEEEmembership{Senior Member,~IEEE}\\
	\IEEEcompsocitemizethanks{
		\IEEEcompsocthanksitem  
        This work was supported in part by the National Natural Science Foundation of China, No.:62576195, and No.:62276155; in part by the Key R\&D Program of Shandong Province (Major scientific and technological innovation projects), China, No.: 2025CXGC020101\\
		Qianyun Yang is with the School of Software, Shandong University, Jinan, 250100, China, and the National Graduate School for Elite Engineers, Shandong University, Jinan, 250100, China. Email: qianyunyang@mail.sdu.edu.cn;
        Zhiwei Chen,~Yupeng Hu,~Zixu Li,~Zhiheng Fu are with the School of Software, Shandong University, Jinan, 250100, China. Email: \{zivczw, lizixu.cs, fuzhiheng8\}@gmail.com; huyupeng@sdu.edu.cn; 
        Liqiang Nie is with the School of Computer Science and Technology, Harbin Institute of Technology (Shenzhen), Shenzhen, 518000, China. Email: nieliqiang@gmail.com.
  }
}  

\markboth{IEEE TRANSACTIONS ON KNOWLEDGE AND DATA ENGINEERING, SUBMISSION 2026}%
{Yang \MakeLowercase{\textit{et al.}}: STABLE: Efficient Hybrid Nearest Neighbor Search via  Magnitude-Uniformity and Cardinality-Robustness}

\maketitle
\IEEEpeerreviewmaketitle

\begin{abstract}
Hybrid Approximate Nearest Neighbor Search (Hybrid ANNS) is a foundational search technology for large-scale heterogeneous data and has gained significant attention in both academia and industry. However, current approaches overlook the heterogeneity in data distribution, thus ignoring two major challenges: the \textit{Compatibility Barrier for Similarity Magnitude Heterogeneity} and the \textit{Tolerance Bottleneck to Attribute Cardinality}. To overcome these issues, we propose the robu\underline{S}t he\underline{T}erogeneity-\underline{A}ware hy\underline{B}rid retrieva\underline{L} fram\underline{E}work, STABLE, designed for accurate, efficient, and robust hybrid ANNS under datasets with various distributions. Specifically, we introduce an enhAnced heterogeneoUs semanTic perceptiOn (AUTO) metric to achieve a joint measurement of feature similarity and attribute consistency, addressing similarity magnitude heterogeneity and improving robustness to datasets with various attribute cardinalities. Thereafter, we construct our Heterogeneous sEmantic reLation graPh (HELP) index based on AUTO to organize heterogeneous semantic relations. Finally, we employ a novel Dynamic Heterogeneity Routing method to ensure an efficient search. Extensive experiments on five feature vector benchmarks with various attribute cardinalities demonstrate the superior performance of STABLE.
\end{abstract}


\section{Introduction}
\begin{figure}[h]
	\centering
    	\vspace{-11pt}
	\includegraphics[width=\linewidth]{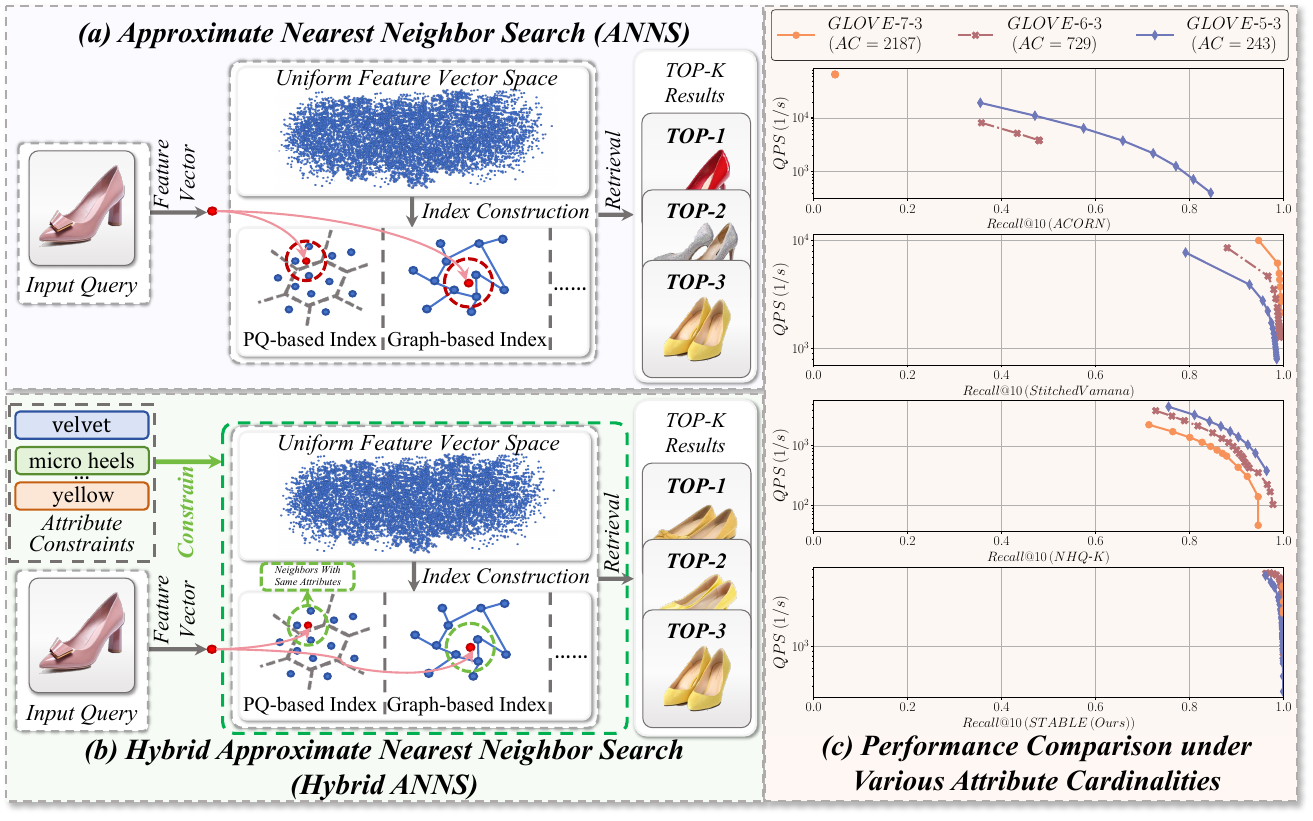}
	\vspace{-18pt}
	\caption{Illustrations of (a) ANNS, (b) Hybrid ANNS, and (c) performance comparison under various attribute cardinalities. }
    \vspace{-11pt}
	\label{fig1}
\end{figure}

\IEEEPARstart{W}{ith} the increasing prevalence of information technology and its applications, vast amounts of unstructured data (e.g., video, image, audio, and text) are generated, shared, and stored on a daily basis~\cite{du2018spell,tekli2016overview,unstructured-datasets,scis-data-2,chen2023bias,Data-to-text-Generation}. Driven by deep representation learning~\cite{Text-search, ahmad2020deep, li2018survey,HINT,MELT}, these large-scale unstructured data can be encoded into corresponding feature vectors and mapped into a unified feature space, enabling retrieval via assessment of feature similarity. Benefiting from conventional Approximate Nearest Neighbor Search (ANNS) methods, researchers can efficiently perform large-scale unstructured data retrieval. Specifically, as shown in Fig.~\ref{fig1}(a), conventional ANNS methods typically construct an index within the unified feature vector space and apply routing algorithms on this index to efficiently retrieve data points that are approximately nearest to the query vector. Accordingly, conventional ANNS methods have been widely adopted across various information-retrieval scenarios~\cite{wang2017reverse,curse-cost,lu2006hierarchical,li2002clustering}, serving as the efficient backbone for tasks such as image/video retrieval, image-text matching, recommendation systems and cross-modal video-moment localization~\cite{hu2021coarse,hu2023semantic,hu2021video}, multimodal retrieval~\cite{retrack,HUD,REFINE}, and multimodal learning~\cite{habit,intent,OFFSET,median,pair,encoder,FineCIR}.

While conventional ANNS methods~\cite{wang2021comprehensive, bruch2023approximate, zuo2023arkgraph,zhao2023towards,yi2016practical,cai2019revisit,Hash2020,ANN-Hd,xu2018online,ozan2016k,wang2014optimized} effectively leverage unstructured data, they cannot support structured attribute information --- labels assigned to the data (manually annotated or extracted from human descriptions) that enable targeted filtering in specific retrieval tasks. This limitation prevents ANNS approaches from supporting personalized retrieval scenarios requiring simultaneous consideration of feature similarity and attribute constraints.
A concrete manifestation of this limitation emerges in e-commerce product retrieval systems. Modern platforms increasingly adopt image-based search paradigms where users query with product photos, employing conventional ANNS to retrieve visually similar items (Fig.~\ref{fig1}(a)). However, users typically want attribute-based filtering~(brand, color, style, material) in addition to visual matching, which is a capability absent in conventional ANNS frameworks.
To overcome this limitation, hybrid ANNS methods have emerged. 
These approaches address this gap by combining efficient retrieval of unstructured features with filtration of structured attributes, thereby imposing fine-grained constraints on the results (Fig.~\ref{fig1}(b)). In large-scale heterogeneous data environments, the hybrid retrieval integrates feature similarity and attribute consistency within the similarity metric, significantly enhancing the personalization and accuracy of the retrieval results. This is a critical capability for high-quality retrieval over large-scale hybrid datasets~\cite{AttributeCon}.

Considering the necessity of hybrid ANNS, several pioneering works~\cite{ADBV,vearch,milvus2021,nhq,acorn,filterDiskann} have been devoted to the \textit{Attribute-Equality} hybrid search scenario. In this scenario, the retrieved items must not only satisfy the similarity with the query in terms of features, but at the same time, their structured attributes must fully match the query's structured attribute requirements. 
For the sake of conciseness, throughout the remainder of this paper, all uses of hybrid ANNS refer specifically to hybrid ANNS targeting the \textit{Attribute-Equality} scenario.
Existing \textit{Attribute-Equality} hybrid ANNS methods can be grouped into three categories: \textit{Pre-filtering}, \textit{Post-filtering}, and \textit{Specialized Indices} for hybrid ANNS.
\textit{Pre-filtering} approaches~\cite{milvus2021,ADBV} first perform attribute filtering and then compute feature similarity to achieve hybrid ANNS. \textit{Post-filtering} approaches~\cite{vearch} reverse this order, computing feature similarity first and then applying attribute filtering to the set of feature-similar vectors.
By contrast, \textit{Specialized Indices} methods~\cite{acorn,nhq,filterDiskann} design a unified index that simultaneously measures feature similarity and attribute consistency, yielding final hybrid ANNS results in a single retrieval process. Although these three categories of methods have demonstrated encouraging performance, they primarily address how to capture the semantics of heterogeneous structured and unstructured data, while lacking attention to the heterogeneity in data distribution. This makes existing methods tend to ignore the following two challenges that negatively impact search performance in the \textit{Attribute-Equality} scenario:

\begin{table}[ht]
  \centering
  \vspace{-8pt}
  \caption{Similarity magnitude statistics across different datasets, where the attribute data and the similarity measurement adhere to the configurations specified in NHQ~\cite{nhq} (i.e., \textcolor{r4}{with a unified attribute vector design, and using} Euclidean distance for feature similarity and Hamming distance for attribute consistency).}
    \vspace{-7pt}
    	\resizebox{0.9\linewidth}{!}{
    \begin{tabular}{l|c|c|c|c|c|c}
      \Xhline{1pt}
		\hline\hline

    \multicolumn{1}{c|}{\multirow{2}{*}{Datasets}} & \multicolumn{3}{c|}{Feature Distance} & \multicolumn{3}{c}{Attribute Distance} \\
\cline{2-7}          & Min   & Max   & Average & Min   & Max   & Average \\
    \hline
    SIFT1M & 8.43  & 719.67 & 536.93 & 0.00  & 3.00  & 1.67 \\
    GLOVE-100 & 1.37  & 22.72 & 7.66  & 0.00  & 3.00  & 1.67 \\
    CRAWL & 1.42  & 16.01 & 7.78  & 0.00  & 3.00  & 1.67 \\
    BIGANN10M & 33.41 & 712.57 & 528.68 & 0.00  & 3.00  & 1.67 \\
    DEEP10M & 0.07  & 1.70  & 1.36  & 0.00  & 3.00  & 1.67 \\
		\hline\hline
  \Xhline{1pt}
    \end{tabular}%
    }
  \vspace{-10pt}
  \label{tab:similarity_h}%
\end{table}%

\textbf{$\bullet$ C1: Compatibility Barrier for Similarity Magnitude Heterogeneity.} 
Existing methods typically apply a \textcolor{r1}{rigid fusion metric that struggles to adjust for diverse real-world datasets, failing to flexibly handle the varying magnitude disparities between feature and attribute spaces}. 
As shown in TABLE~\ref{tab:similarity_h}, we sample $1,000$ vectors from mainstream datasets and report the \textcolor{r1}{statistics}. 
From these results, we observe \textcolor{r1}{a pronounced disparity between feature distance and attribute distance that varies significantly across datasets}. 
For example, in SIFT1M, the average feature distance \textcolor{r1}{($536.93$) is hundreds of times larger than} the average attribute distance \textcolor{r1}{($1.67$)}. 
Conversely, in DEEP10M, the average feature distance \textcolor{r1}{($1.36$) is comparable to the attribute distance ($1.67$).} 
\textcolor{r1}{Critically, such magnitude disparities are difficult to resolve via standard scaling, as hybrid retrieval necessitates a balanced joint evaluation of feature similarity and attribute matching, where improper scaling risks disrupting the ranking logic by causing one signal to dominate. Furthermore, identifying a unified fusion strategy that remains robust across diverse data distributions is non-trivial. Consequently, the lack of a high-quality metric to adaptively accommodate data diversity constitutes a critical bottleneck constraining the robustness of current hybrid ANNS.}


\textbf{$\bullet$ C2:~Tolerance Bottleneck to Attribute Cardinality.} \textcolor{r1}{This refers to the robustness of retrieval performance against variations in attribute diversity.} As illustrated in Fig.~\ref{fig1}(c), the search performance of the state-of-the-art methods ACORN, StitchedVamana, and NHQ degrades markedly as attribute cardinality (i.e., the number of distinct attributes in the dataset) changes, whereas real-world scenarios often involve attribute updates. Consequently, developing index architectures that robustly adapt to various attribute cardinalities remains a critical challenge.

To address these challenges, we propose a robu\textbf{S}t he\textbf{T}erogeneity-\textbf{A}ware hy\textbf{B}rid retrieva\textbf{L} fram\textbf{E}work, named STABLE. STABLE constructs a graph index via the \textit{enh\textbf{A}nced heterogeneo\textbf{U}s seman\textbf{T}ic percepti\textbf{O}n (AUTO)} metric and couples it with \textit{Dynamic Heterogeneity Routing} to achieve efficient and robust hybrid ANNS. 
Specifically, we first design the AUTO metric to redefine distance measurements during graph-index construction. 
By incorporating dataset statistics information, the metric reconciles similarity magnitude heterogeneity between feature similarity and attribute consistency within a dataset, enabling precise evaluation of heterogeneous semantics. That is, it ensures that data nodes similar in both feature and attribute spaces are drawn closer, while dissimilar nodes are effectively pushed apart. Moreover, the metric can also be automatically adapted to different datasets, yielding increased robustness to data heterogeneity.
Then, we design our \textit{\textbf{H}eterogeneous s\textbf{E}mantic re\textbf{L}ation gra\textbf{P}h (HELP)} index, innovatively introducing a heterogeneous semantic pruning mechanism. After organizing heterogeneous semantic relations by connecting semantically similar data nodes, HELP can minimize redundant connections with the help of the heterogeneous semantic pruning mechanism to support efficient routing afterward. 
Finally, we introduce \textit{Dynamic Heterogeneity Routing}, a new and fast routing strategy, ensuring the search efficiency.
Extensive experiments on datasets with various attribute cardinalities demonstrate that STABLE outperforms existing methods in both search performance and robustness.

Overall, the main contributions of this paper are as follows:

\begin{itemize}
    \item We propose the AUTO metric, the first hybrid ANNS metric to enable robust data semantic similarity modeling by introducing dataset statistics for automatic adjustment. It can reconcile magnitude differences between feature similarity and attribute consistency in a unified metric, while robustly coping with diverse datasets.
    \item We introduce a heterogeneous semantic pruning mechanism to build our HELP index. This index is used to efficiently establish heterogeneous semantic connections between similar data nodes based on AUTO. Combined with a novel routing strategy named \textit{Dynamic Heterogeneity Routing}, the efficiency of hybrid ANNS can be further improved.
    \item We conduct extensive experiments on datasets derived from five feature-vector datasets with various attribute cardinalities, validating STABLE's superiority and robustness across multiple dimensions, including search accuracy, computational efficiency, and robustness. To our knowledge, we are the first to systematically investigate search robustness from the perspective of variations of attribute cardinality.
\end{itemize}

\section{RELATED WORK}\label{sec2}
\subsection{Conventional ANNS}

Conventional Approximate Nearest Neighbor Search (ANNS)~\cite{ANN1998} methods aim to achieve high search accuracy while maintaining desirable search efficiency, thus enabling high-quality search for large-scale data. 
Specifically, they aim to utilize feature similarity calculation to find the target data corresponding to the given query, which can be further divided into $4$ subcategories: hash-based, tree-based, quantization-based, and graph-based ANNS methods. 

Concretely, hash-based methods~\cite{Hash2015,ANN-Lazylsh,Hash2020,Hash2020-1} first encode the given feature vectors into compact hash codes, while maintaining the semantic relevance between them, and then store them in the corresponding hash table. Thereafter, for a given query, the top-k nearest neighbors can be obtained via Hamming distance measurement. 
Compared with hash-based methods, the tree-based methods~\cite{ANN-Hd,Tree2014} allow data to be stored in a tree index without encoding transformation, and search can be performed based on semantic pruning of the tree index. Moreover, the quantization-based methods~\cite{Quan2011,Quan2014,Quan2015,Quan2020} first perform dimension reduction to project the original data into corresponding quantified space, then perform simplified distance computation to filter out the data candidates, and finally get the target top-k nearest neighbors by complete feature distance calculation~\cite{ANN-Hd,b-tree,Tree2008,Tree2014}.

Apparently, these methods still have to perform a major traversal search from the corresponding hash tables, index trees, and quantified space to find the target data. Namely, they lack an effective representation of the semantic similarity between nearest neighbors, thus failing to further improve search performance. Therefore, to accurately represent the semantic relationships between nearest neighbors, the graph-based ANNS methods have been proposed~\cite{NSG,SSG,DiskAnn,HNSW2020,FANNG}. Following the idea of ``neighboring nodes may also be neighbors'', these methods use the connection-pruning strategy to construct the corresponding graph index, which can provide a more efficient representation of the near-neighborhood relationship than non-graph-based methods. Therefore, based on the approximate connectivity of the graph index, the effective ANNS can be achieved completely. Available research evidence shows that these graph-based methods perform better than the other three types of methods in Ann-benchmark datasets~\cite{ann-bench}. 

Although the four method types have achieved certain retrieval efficacy, they can only perform feature similarity computation based ANNS, and ignore the necessity of comprehensive evaluation between feature similarity and attribute equality, thus failing to complete hybrid ANNS under collaborative constraints on feature similarity and attribute consistency.

\subsection{\textcolor{r2}{Hybrid ANNS}}

\textcolor{r2}{Hybrid ANNS, also referred to \textit{Filtered Approximate Nearest Neighbor Search (FANNS)}, aims to retrieve nearest neighbors from a feature vector set subject to scalar attribute constraints. Following the pruning-focused framework proposed in recent surveys~\cite{lin2025survey,chronis2025filtered}, we classify existing methods into two macro-categories: Disjoint Pruning (VSP \& SSP) and Joint Pruning (SJP \& VJP).}

\noindent \textit{\textbf{\textcolor{r2}{1) Disjoint Pruning Strategies (VSP \& SSP).}}} 
\textcolor{r2}{These methods treat feature vector search and attribute filtering as two independent stages.}

\textit{\textcolor{r2}{\underline{Scalar-Solely Pruning (SSP):}}} \textcolor{r2}{Also known as \textit{Pre-filtering}. Methods like ADBV~\cite{ADBV} and standard Milvus~\cite{milvus2021} first identify a subset of data satisfying the attribute constraint (scalar pruning) and then perform a vector search (typically brute-force or quantization-based) within this subset. While effective for highly selective queries (small subsets), SSP suffers from high computational costs when the filtered subset remains large, as it fails to leverage vector indexing for pruning.}

\textit{\textcolor{r2}{\underline{Vector-Solely Pruning (VSP):}}} \textcolor{r2}{Also known as \textit{Post-filtering}. Methods like Vearch~\cite{vearch} first perform a standard ANNS to retrieve top-$K'$ candidates (vector pruning) and then filter out those mismatching the attributes. The core challenge of VSP lies in estimating the optimal search depth $K'$. An insufficient $K'$ leads to fewer than $K$ valid results (low recall), while an excessively large $K'$ degrades efficiency.}

\noindent \textit{\textbf{\textcolor{r2}{2) Joint Pruning Strategies (SJP \& VJP).}}}
\textcolor{r2}{To overcome the limitations of disjoint strategies, recent works propose joint pruning to leverage both feature vector and scalar attribute information simultaneously.}

\textit{\textcolor{r2}{\underline{Scalar-Centric Joint Pruning (SJP):}}} 
\textcolor{r2}{Methods in this category organize indices based on scalar distributions. 
    HQI~\cite{HQI} physically partitions the dataset into subsets.
    MA-NSW~\cite{ma-nsw} constructs multiple NSW graphs for different attribute values to accelerate filtering.
    UNG~\cite{ung} further connects these subgraphs based on containment relationships.
    While effective for simple filters, SJP methods like MA-NSW suffer from prohibitive memory costs when handling complex attributes with high cardinality, as maintaining indices for all attribute combinations is impractical.}

    \textit{\textcolor{r2}{\underline{Vector-Centric Joint Pruning (VJP):}}} \textcolor{r2}{This category integrates scalar filtering into the vector index traversal.
    HQANN~\cite{hqann} and NHQ~\cite{nhq} map attributes and features into a unified ``fusion space'' for indexing.
    Filtered-DiskANN~\cite{filterDiskann} constructs graphs by connecting nodes sharing attributes.
    ACORN~\cite{acorn} modifies HNSW to perform predicate-subgraph traversal.
    However, these methods face robustness challenges. Metric fusion methods (HQANN, NHQ) often fail to address Magnitude Heterogeneity, while subgraph-based methods (ACORN) degrade under varying \textit{Attribute Cardinality}.}

\underline{\textit{\textbf{\textcolor{r2}{Gaps of Existing Methods.}}}}
\textcolor{r2}{Although VJP methods like ACORN and NHQ have advanced the field, they exhibit two critical limitations regarding robustness:
(1) Existing metric fusion strategies (e.g., in NHQ) often struggle to adjust for diverse real-world datasets, failing to flexibly handle the varying magnitude disparities between feature and attribute spaces.
(2) Existing methods like ACORN and StitchedVamana~(variant of Filtered-DiskANN) suffer significant performance degradation when the attribute cardinality varies. The connectivity of its index deteriorates significantly under high cardinality settings.
In summary, the performance bottleneck caused by insufficient robustness remains an unresolved challenge and is precisely the issue that our STABLE addresses in this paper.
}

\section{METHODOLOGY}\label{sec3}


\textcolor{r2}{As the primary innovation, our proposed STABLE is designed to achieve efficient and robust hybrid ANNS through three synergized algorithmic contributions tailored for heterogeneous data distributions. Specifically: 
(1) the \textit{enh\textbf{A}nced heterogeneo\textbf{U}s seman\textbf{T}ic percepti\textbf{O}n (AUTO)} metric introduces a statistically calibrated measurement that adaptively balances feature similarity and attribute consistency to robustly resolve similarity magnitude heterogeneity; 
(2) the \textit{\textbf{H}eterogeneous s\textbf{E}mantic re\textbf{L}ation gra\textbf{P}h (HELP)} index implements a novel heterogeneous semantic pruning mechanism to construct topological highways across different attribute subspaces, thereby organizing heterogeneous semantic relations for efficient navigation; 
and (3) the \textit{Dynamic Heterogeneity Routing} adopts a novel coarse-to-fine strategy utilizing a dynamic pioneer set, which allows the search process to effectively escape the local optima inherent in heterogeneous spaces for accurate and efficient retrieval.}


In this section, we first formulate the hybrid ANNS task (detailed in Section~\ref {Problem Formulation}). Subsequently, we elaborate on our proposed AUTO metric (described in Section~\ref {AUTO}), HELP index (presented in Section~\ref {sec:HELP}), and the dynamic heterogeneity routing (explained in Section~\ref {Local Routing Process}). 
\textcolor{r3}{For ease of understanding, we summarize the key notations that appear in the following TABLE~\ref{tab:notations}:}

\begin{table}[h]
\caption{Summary of Key Notations}
\label{tab:notations}
\centering
  	\resizebox{\linewidth}{!}{
\begin{tabular}{l|l}

\Xhline{1pt}
\hline\hline
\textbf{Symbol} & \textbf{Description} \\
\rowcolor[rgb]{ .949,  .949,  .949}
\multicolumn{2}{c}{\textit{\textbf{Data \& Problem Definition}}} \\

$\mathcal{D}$ & The dataset containing $N$ data nodes, $\mathcal{D}=\{D_1, ..., D_N\}$ \\
$N$ & Total number of data nodes in the dataset \\
$D_i$ & The $i$-th data node, consisting of feature and attribute vectors \\
$V_i, A_i$ & The feature vector and attribute vector of node $D_i$ \\
$\mathcal{Q}$ & The query node, consisting of $\hat{V}$ and $\hat{A}$ \\
$\hat{V}, \hat{A}$ & The feature vector and attribute vector of the query \\
$M, L$ & Dimensions of the feature vector and attribute vector \\
$K$ & The number of nearest neighbors to retrieve (Top-K) \\
$\Theta$ & Attribute cardinality (number of distinct attribute values) \\

\rowcolor[rgb]{ .949,  .949,  .949}
\multicolumn{2}{c}{\textit{\textbf{AUTO Metric \& HELP Index}}} \\
$\mathcal{S}_V(\cdot)$ & Feature similarity measurement function (Euclidean distance) \\
$\mathcal{S}_A(\cdot)$ & Attribute consistency measurement function (Manhattan distance) \\
$\alpha$ & Adaptive trade-off parameter in the AUTO metric \\
$\mathcal{U}(\cdot)$ & The proposed AUTO metric for heterogeneous similarity \\
$\overline{\mathcal{S}_V}, \overline{\mathcal{S}_A}$ & Average feature/attribute distance of sampled nodes \\
$\mathcal{I}$ & The HELP graph index structure, $\mathcal{I}=(D, E)$ \\
$E$ & The set of edges in the graph index \\
$\Gamma$ & Maximum number of neighbors for each node \\
$\mathcal{N}_i$ & The neighbor set of node $D_i$ \\
$\Gamma_{new}$ & Maximum number of newly added neighbors in iteration \\
$\sigma$ & Cosine similarity threshold for heterogeneous semantic pruning \\
\rowcolor[rgb]{ .949,  .949,  .949}
\multicolumn{2}{c}{\textit{\textbf{Dynamic Heterogeneity Routing}}} \\

$\mathcal{R}$ & The result set (candidate pool) maintained during search \\
$\mathcal{P}$ & The Pioneer Set used for dynamic coarse routing \\
$P$ & The size of the Pioneer Set \\
$l_{\mathcal{P}}$ & Path length effectively skipped by the Pioneer Set \\

\hline\hline
\Xhline{1pt}
\end{tabular}
}
\end{table}

\subsection{Preliminaries}
\label{Problem Formulation}
Following previous works~\cite{nhq,filterDiskann}, STABLE aims to address the challenging hybrid ANNS task on the \textit{Attribute-Equality} scenario. The goal of this scenario is to efficiently search the data nodes that are most similar to the query, which necessitates the concurrent satisfaction of both proximity in feature vector values and exact congruence in attribute values.
Given a data node set, named as $\mathcal{D}$ containing $N$ nodes, formulated as $\mathcal{D}=\{D_1,\dots,D_i,\dots,D_N\},$
where $1\leq i\leq N$, and $D_i$ denotes a data node including feature vector $V_i$ and attribute vector $A_i$, expressed as $D_i=\{V_i, A_i\}$. 
Assume that there is a given query node with its own feature and attribute vectors, denoted as $\mathcal{Q}=\{\widehat{V}, \widehat{A}\}$. We expect to design a hybrid ANNS framework $\mathcal{F}$ to effectively retrieve the target node list $\mathcal{T}$ from $\mathcal{D}$, whose attributes fully match the constraints of $\widehat{A}$. The process is formulated as $\mathcal{T} \Leftarrow \mathcal{F}\left(\mathcal{D},\mathcal{Q}  \right),$
where $\mathcal{T}$ is the top-$K$ list containing $K$ target nodes that are most similar to $\mathcal{Q}$. In the attribute equality scenario, the ``fully match'' represents $A_i=\hat{A}$.
And without loss of generality, assume that the dimension of $V_i$ and $\widehat{V}$ is $M$, and the dimension of $A_i$ and $\widehat{A}$ is $L$, respectively. 

\subsection{Enhanced Heterogeneous Semantic Perception (AUTO)} 
\label{AUTO}
To address the similarity magnitude heterogeneity, we propose an \textit{enh\textbf{A}nced heterogeneo\textbf{U}s seman\textbf{T}ic percepti\textbf{O}n (AUTO)} metric. This metric is designed to redefine the distance metric for collaboratively measuring feature similarity and attribute consistency. In this part, we first describe the numerical mapping introduced for the attribute vectors, and then detail the AUTO metric and its correctness.

\subsubsection{Attribute Vector Numerical Mapping} 
\label{Attribute Vector Numerical Mapping}
To make AUTO adaptable to diverse attribute vector types (e.g., numerical and categorical), we employ a numerical mapping process to convert each attribute vector into a numerical vector that is more convenient for computation.

Specifically, let $\mathcal{A}_l=\{\textbf{a}_1,..., \textbf{a}_u,...,\textbf{a}_{U_l}\}$ denote the attribute value set for the $l$-th dimension of the attribute vectors, where $U_l$ represents the attribute cardinality of the value set in the $l$-th dimension. 
Subsequently, assuming that $a_l^i$ denotes the $l$-th dimension of attribute vector $A_i$, for $a_l^i=\textbf{a}_u$, we define the following numerical mapping function to map the $a_l^i$ to a numerical value, facilitating subsequent calculations of the AUTO metric, formulated as follows,
\begin{equation} 
	\label{map}
	a_l^i=\operatorname{MAP}(\textbf{a}_u)=u.
\end{equation}
That is, we map the attribute value of the \(l\)-th dimension of the attribute vector \(A_i\) to the position \textit{ID} where it appears in the attribute value set \(\mathcal{A}_l\). Note that after the numerical mapping process, the attribute vectors can still be used to evaluate attribute consistency, as stated in Remark~\ref{lemma1}.
\begin{Remark}
\label{lemma1}
    The attribute consistency evaluation aims to identify nodes that fully match the attribute constraints, and the proposed numerical mapping does not affect the results of the full matching check. That is, if~$\{a_l^i\!\!=\!\!\textbf{a}_u\}\!\!\!\overset{\text{\textit{not match}}}\Longleftrightarrow \!\!\!\{a^{j}_l\!\!=\!\!\textbf{a}_{u^{\prime}}\}$, then $\{a_l^i\!\!=\!\!\operatorname{MAP}(\textbf{a}_u)\!\!=\!\!u\}\!\!\!\overset{\text{\textit{not match}}}\Longleftrightarrow \!\!\!\{a^{j}_l\!\!=\!\!\operatorname{MAP}(\textbf{a}_{u^{\prime}})\!\!=\!\!u^{\prime}\}$, and vice versa.
\end{Remark}
For ease of representation, the values of the attribute vectors mentioned in the subsequent contents are the values after performing the numerical mapping.




\subsubsection{Calculation of the AUTO Metric} 

Conventional ANNS indices evaluate semantic similarity using distance metrics oriented only towards feature vectors. With the introduction of attribute vectors, these distance metrics are difficult to accommodate the similarity magnitude heterogeneity between feature distance and attribute distance.
To address this limitation, we design a new fusion metric, abbreviated as the AUTO metric, which effectively fuses feature similarity and attribute consistency to measure the similarity between two data nodes.

As aforementioned, one of the purposes of hybrid ANNS is to retrieve results that fully match the attribute constraints of the input query. For this reason, when constructing the index, we desire that when the attribute vectors of two nodes are different, these two nodes ought to be treated as dissimilar nodes. Hence, the distance between them in the graph ought to be as large as possible to better distinguish nodes with different attribute constraints.
Based on this intuition, the AUTO metric aims to evaluate the heterogeneous semantic similarity among each $D_i\!\!=\!\!\{V_i, A_i\}\!\!\in\!\!\mathcal{D}$ and the given query $\mathcal{Q}\!\!=\!\!\{\widehat{V},\widehat{A}\}$ to push nodes with different attribute vectors as far away from each other as possible.
In the following, we take the heterogeneous semantic similarity between node $D_i$ and the query $\mathcal{Q}$ as an example to introduce the calculation process of AUTO.

Specifically, the heterogeneous semantic similarity evaluation consists of two basic measurement metrics: \textit{attribute consistency evaluation} and \textit{feature similarity calculation}, formulated as follows.

\noindent
\textbf{$\bullet$~\underline{Attribute consistency evaluation}.} With the numerical mapping (depicted in Section~\ref{Attribute Vector Numerical Mapping}), we are capable of quantitatively evaluating attribute consistency. We apply the Manhattan distance to efficiently push away nodes with different attribute constraints. Specifically, the attribute consistency between $A_i$ and $\widehat{A}$ is estimated as follows,
\begin{equation}
    \label{ad}
    \mathcal{S}_A(A_i,\widehat{A}) = \sum_{l=1}^{L}|a^i_l - \widehat{a}_l|,
\end{equation}
where $a^i_l$ and $\widehat{a}_l$ separately denote the value of the $l$-th dimension of attribute vectors $A_i$ and $\widehat{A}$.

\textcolor{r1}{We employ the Manhattan distance for dual strategic benefits. 
First, regarding the potential ordinal bias for nominal attributes, we leverage it as a functional advantage rather than a drawback. As analyzed in the subsequent Remark~\ref{reason Manhattan}, the cumulative distance generated by mismatches functions as a ``soft barrier,'' which effectively amplifies the distinguishability of non-compliant nodes and enhances filtering efficiency during graph routing. Second, unlike the Hamming distance, which is limited to equality checks, this metric naturally accommodates magnitude variations, allowing our framework to be easily extended to support ordinal attributes (e.g., ratings, years) in complex real-world scenarios.}


\begin{Remark}
\label{reason Manhattan}
The Manhattan distance, compared to the Euclidean and Hamming distances, more robustly and effectively distinguishes nodes with different attributes.
\end{Remark}
\textit{\underline{Explanation.} Let $D_i=\{V_i, A_i\}$ and $D_j=\{V_j, A_j\}$ denote two nodes with distinct attribute vectors in $\mathcal{D}$. 
From Eq.~\ref{ad}, the attribute consistency via Manhattan distance is $\mathcal{S}_A(A_i,A_j) = \sum_{l=1}^{L}|a^i_l - a^j_l|$. 
For comparison, the Euclidean distance is $\mathcal{S}_A^{E}(A_i,A_j)=\sqrt{\sum_{l=1}^{L}(a^i_l - a^j_l)^2}$, and the Hamming distance is $\mathcal{S}_A^{H}(A_i,A_j)=\sum_{l=1}^{L}\mathbb{I}(a^i_l \neq a^j_l)$, where $\mathbb{I}(\cdot)$ is the indicator function (equal to 1 if $a^i_l \neq a^j_l$, and 0 otherwise). 
Since the attribute values are mapped to integers, any mismatch implies $|a^i_l - a^j_l| \ge 1$. 
Consequently, we derive the inequality relationships: $\mathcal{S}_A(A_i,A_j) \geq \mathcal{S}_A^{E}(A_i,A_j) \geq 1$ and $\mathcal{S}_A(A_i,A_j) \geq \mathcal{S}_A^{H}(A_i,A_j) \geq 1$. 
This demonstrates that when attribute constraints differ, the Manhattan distance consistently yields a larger separation value than other metrics, thereby achieving a more effective distinction between nodes with different attribute constraints.
}

 \noindent
\textbf{$\bullet$~\underline{Feature similarity calculation}.} Following previous ANNS methods~\cite{SSG,nsw}, the similarity between the feature vectors $V_i$ and $\widehat{V}$ is measured via Euclidean distance, formulated as,
	\begin{equation}
		\label{vd}
		\mathcal{S}_V(V_i,\widehat{V}) = \sqrt{\sum_{m=1}^{M}(v^i_m-\widehat{v}_m)^2},
	\end{equation}
	where $v^i_m$ and $\widehat{v}_m$ separately denote the values of the $m$-th dimension of feature vectors $V_i$ and $\widehat{V}$.

Based on the above two measurement metrics, we design the AUTO metric to enable a high-quality collaborative measurement of feature similarity and attribute consistency. As shown in Table~\ref{tab:similarity_h}, due to the presence of similarity magnitude heterogeneity, rather than simply summing \textbf{Feature similarity} and \textbf{Attribute consistency}, we utilize the \textbf{Attribute consistency} as a weight on the \textbf{Feature similarity} for the AUTO metric. This operation balances the magnitudes of $\mathcal{S}_A(A_i,\widehat{A})$ and $\mathcal{S}_V(V_i,\widehat{V})$, preventing one metric from being neglected when the other's value is substantially larger. 
Concretely, the overall similarity evaluation $\mathcal{U}(D_i,\mathcal{Q})$ of the AUTO metric between each $D_i \in \mathcal{D}$ and the given query $\mathcal{Q}$ can be formulated as,
\begin{equation}
	\label{fd}
	\mathcal{U}(D_i,\mathcal{Q})=\mathcal{S}_V(V_i,\widehat{V})\times(1+\dfrac{\mathcal{S}_A(A_i,\widehat{A})}{\alpha}),
\end{equation}
where $\alpha$ denotes the trade-off parameter between these two measurement metrics, which is defined to accommodate data with diverse distributions and enhance the formula's compatibility with similarity magnitude heterogeneity. Notably, the smaller value of $\mathcal{U}(D_i,\mathcal{Q})$ indicates greater similarity between $D_i$ and $\mathcal{Q}$. 

To determine an appropriate value for $\alpha$, we first randomly sampled $1,000$ nodes from the dataset prior to index construction and computed the average feature distance $\overline{\mathcal{S}_V}$ and the average attribute distance $\overline{\mathcal{S}_A}$. We then used these two average values as a reference to set the value of $\alpha$, formulated as,
\begin{equation}
	\label{alp}
	\alpha=\operatorname{Norm}( \dfrac{N}{\overline{\mathcal{S}_V}})+ \operatorname{Norm}(\dfrac{\overline{\mathcal{S}_A}}{L}),
\end{equation}

where \(N\) denotes the total number of nodes in the dataset, and \(\operatorname{Norm}(\cdot)\) is our defined linear scaling function, which repeatedly multiplies or divides its input by $10$ to map the value into the interval \((0.1, 1]\). Intuitively, a larger \(N\) makes it harder to retrieve feature\mbox{-}similar nodes in the datasets, while a smaller average feature distance \(\overline{\mathcal{S}_V}\) implies that feature similarity between node pairs is harder to discriminate. 
Consequently, an increase ratios $\frac{N}{\overline{\mathcal{S}_V}}$ signifies greater difficulty in feature\mbox{-}similarity discrimination. In such cases, the weight on feature distance in the AUTO metric should be increased accordingly to better discriminate the feature\mbox{-}similarity, while the weight on attribute consistency should be relatively decreased (i.e., \(\alpha\) should grow). The same rationale applies when analyzing $\frac{\overline{\mathcal{S}_A}}{L}$.
Thus, by balancing the weights of feature distance and attribute consistency, the proposed AUTO metric adaptively accommodates similarity magnitude heterogeneity, yielding improved retrieval performance. The validation of the $\alpha$ calculation is presented in Section~\ref{appendix:alpha}.


\vspace{0.5em}
\subsubsection{\textcolor{r2}{Rationale Analysis and Justification}}
\textcolor{r2}{To rigorously justify the design of our AUTO metric, we provide a comprehensive analysis covering its design rationale, formal correctness, mathematical validity, and comparative advantages over existing techniques.}

\vspace{0.3em}
\noindent \textit{\textbf{\textcolor{r2}{[a] Rationale of Soft Navigation for Hard Constraints.}}}
\textcolor{r2}{A distinctive design of STABLE is the transformation of discrete attribute constraints into a similarity metric, which can then be utilized as penalty terms in Eq.~\ref{fd}. While user intent typically requires exact attribute matching, enforcing this via ``hard filtering'' often severs navigational paths, creating isolated subgraphs (i.e., the Connectivity Barrier). By incorporating attribute consistency as a weighted term, we convert the rigid Boolean constraint into a ``Soft Navigation'' mechanism. This allows the algorithm to temporarily traverse nodes with mismatched attributes—acting as bridges—to escape local optima. Crucially, the introduced penalty term ensures that the distance for attribute-matching nodes remains the original feature distance, whereas the distance for mismatched nodes is amplified multiplicatively. Consequently, although the process employs ``soft'' paths to achieve connectivity, the final results effectively converge to the ``hard'' exact-match targets.}

\vspace{0.3em}
\noindent \textit{\textbf{\textcolor{r4}{[b] Formal Proof of Selection Correctness.}}}
\textcolor{r4}{To explore in depth how the metric selects Top-K results, we formally prove its correctness in handling nodes with mismatched attributes. Let $D_{match}$ be a node with a perfect attribute match ($S_A=0$) and feature distance $S_V^{match}$. Let $D_{mism}$ be a node with an attribute mismatch ($S_A \ge 1$) and feature distance $S_V^{mism}$. The AUTO metric ranks $D_{mism}$ better than $D_{match}$ (i.e., $\mathcal{U}(D_{mism}) < \mathcal{U}(D_{match})$) if and only if:}
\begin{equation}
\label{formal}
\textcolor{r4}{S_V^{mism} \times (1 + \frac{S_A}{\alpha}) < S_V^{match} \iff S_V^{mism} < \frac{S_V^{match}}{1 + \lambda}}
\end{equation}
\textcolor{r4}{where $\lambda = S_A/\alpha$ serves as the penalty factor. This inequality proves that the selection condition is independent of the absolute scale of feature distances. Instead, it imposes a \textit{relative margin}: a mismatched node is selected only if its feature similarity is significantly stronger (by a factor of $1+\lambda$) than that of a matching node. This allows us to preserve ``bridge nodes'' for connectivity while excluding other nodes with mismatched attributes due to the large, $S_A$-proportional~$\lambda$.}

\vspace{0.3em}
\noindent \textit{\textbf{\textcolor{r1}{[c] Geometric Validity and Triangle Inequality.}}}
\textcolor{r1}{We further analyze the validity of the AUTO metric through its dual geometric properties:}
\begin{itemize}[leftmargin=12pt]
    \item \textcolor{r1}{\textbf{Strict Metricity in Attribute-Uniform Contexts:} Within any subspace sharing consistent attributes (e.g., $\mathcal{S}_A = c$), the AUTO metric strictly reduces to a linearly scaled Euclidean distance: $\mathcal{U} = (1 + c/\alpha) \cdot \mathcal{S}_V$. As $\mathcal{S}_V$ satisfies the triangle inequality, this linear scaling preserves the rigorous geometric property locally, guaranteeing search reliability within attribute clusters.}
    \item \textcolor{r1}{\textbf{Controlled Relaxation for Global Navigability:} Globally, the term $(1 + \mathcal{S}_A/\alpha)$ introduces a necessary ``Triangle Inequality Relaxation.'' This mathematically validates the ``Bridge Nodes'' concept mentioned in the Rationale, preventing the graph from fragmenting into isolated ``Data Islands'' and enabling effective traversal across subspaces with differing attribute values.}
\end{itemize}

\vspace{0.3em}
\noindent \textit{\textbf{\textcolor{r1}{[d] Analytical Derivation of $\alpha$.}}}
\textcolor{r1}{We clarify that $\alpha$ is not a heuristic constant but a Scale Alignment Coefficient derived from the statistical properties of the two heterogeneous spaces to normalize the ``discrimination difficulty'' ($\delta_V$) and ``magnitude scale'' ($\delta_A$).}
\textcolor{r1}{Specifically, $\delta_V \propto N / \bar{\mathcal{S}}_V$ quantifies the density of the feature space, where a crowded space requires a larger $\alpha$ to prevent attribute penalties from masking subtle feature differences. Conversely, $\delta_A \propto \bar{\mathcal{S}}_A / L$ anchors the penalty to the intrinsic scale of the attribute space. Together, these ensure $\alpha$ is an objective, dataset-dependent factor that statistically balances heterogeneous contributions.}

\vspace{0.3em}
\noindent \textit{\textbf{\textcolor{r2}{[e] Qualitative Comparison and Robustness.}}}
\textcolor{r2}{Finally, we justify our design choice by comparing AUTO with prior paradigms:}
\begin{itemize}[leftmargin=12pt]
    \item \textcolor{r2}{\textbf{Vs. Static Linear Metrics (e.g., $\mathcal{S}_{V} + \mathcal{S}_{A}$):} As proven in Eq.~\ref{formal}, simple summation fails when $\mathcal{S}_{V}$ numerically dominates $\mathcal{S}_{A}$ (e.g., $S_V \approx 500$ vs. $S_A \approx 1$), effectively drowning out the attribute constraint. AUTO's multiplicative structure acts as a dynamic scaling factor, ensuring attribute mismatches amplify the total distance \textit{proportionally} regardless of the absolute scale.}
    
    \item \textcolor{r2}{\textbf{Vs. Standard Normalization (e.g., Min-Max):}} \textcolor{r2}{Standard normalization maps metrics to a common scale based on extreme values ($d_{min}, d_{max}$). However, in high-dimensional vector spaces, distance distributions are often skewed or contain outliers. \textcolor{r3}{This limitation is exacerbated in real-world datasets with non-uniform attribute distributions or extreme values, where relying on extreme values causes Min-Max to compress the effective distinguishing range of nearest neighbors into a negligible interval.} In contrast, AUTO utilizes dataset statistics (i.e., average distances $\overline{\mathcal{S}_{V}}$ and $\overline{\mathcal{S}_{A}}$) to calibrate $\alpha$. This design is statistically robust to outliers and adapts to non-uniform distributions, preserving high discriminative resolution in the dense regions where target neighbors are located.}
\end{itemize}

\subsection{Heterogeneous Semantic Relation Graph~(HELP)}
\label{sec:HELP}



\begin{figure*}[ht]
 \begin{center}
     \includegraphics[width=\linewidth]{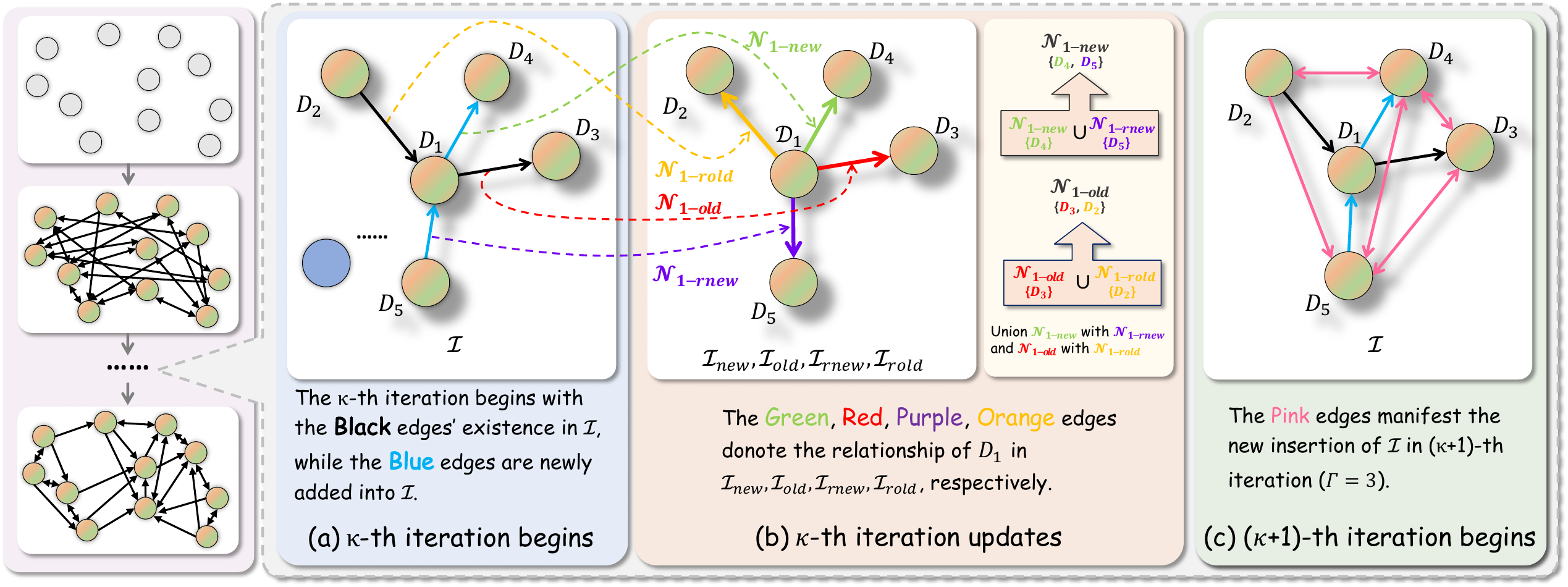}
 \end{center}
 \vspace{-12pt}
	\caption{The iteration of HELP construction. The figure illustrates the neighbor update of $D_1$, with the blue node indicating the other nodes within $\mathcal{D}$.}
	\label{fig:HELP-Index}
 \vspace{-20pt}
\end{figure*}

With the assistance of the AUTO metric, we can obtain the top-$K$ nodes for a given query through traversal search. However, the brute-force search can only achieve sub-optimal retrieval performance on large-scale datasets. Therefore, we construct the \textit{\textbf{H}eterogeneous s\textbf{E}mantic re\textbf{L}ation gra\textbf{P}h (HELP)} index based on the AUTO metric to organize heterogeneous semantic relations for effective hybrid ANNS.


\textcolor{r4}{Before delving into algorithmic details, we outline a few general principles that govern the construction of our index and the rationale behind edge retention. Our overarching goal is to build a navigable graph by balancing two complementary objectives:
(1) Geometric Proximity: We prioritize retaining edges between nodes with high similarity. These connections ensure that the greedy search can find accurate targets.
(2) Semantic Connectivity: To prevent the graph from fracturing into isolated ``Attribute Islands,'' we apply special treatments to preserve heterogeneous ``bridges''---edges linking nodes with high feature similarity but mismatched attributes. These bridges enable the router to traverse across attribute boundaries and escape local optima.
Consequently, while we explicitly protect these critical heterogeneous connections, edges that are geometrically redundant and offer no unique navigational gain are discarded. Guided by these principles, our construction algorithm and the heterogeneous semantic pruning mechanism are designed to operationalize this balance.}

The overall HELP construction is summarized in Algorithm~\ref{HELP-Iteration}, which contains four main steps as follows.

\textbf{\textit{(1) Initialization.}} First, we initialize the index $\mathcal{I}$ whose data node set $D=\mathcal{D}$ and edge set $E=\varnothing$ and randomly generate $\varGamma$ neighbors for each node within $\mathcal{D}$ (\textit{Line 1-5} in Algorithm~\ref{HELP-Iteration}). The edge $\overrightarrow{D_aD_b}$ is created when $D_b$ is a neighbor of $D_a$.

\textbf{\textit{(2) Iteratively connect nodes with approximate semantics.}} Subsequently, we utilize the concept of \textit{reverse neighbors} to iteratively update the neighbors in the HELP index, i.e., \textit{``if $D_i$ is a neighbor of $D_j$, then $D_j$ is a reverse neighbor of $D_i$''}.
Initially, we define $\mathcal{I}$ as the final index to be obtained. $\mathcal{I}_{new}$ and $\mathcal{I}_{rnew}$ are subindexes used to assist in the updating of $\mathcal{I}$. These two subindexes contain the new-added neighbor relationship and the new-added reverse neighbor relationship of each data node respectively. Similarly, the $\mathcal{I}_{old}$ and $\mathcal{I}_{rold}$ are denoted as the subindexes that contain the old neighbor relationship and old reverse neighbor relationship, respectively. Assume that the neighbor sets of each node $D_i\in D$ in $\mathcal{I}_{new}$ and $\mathcal{I}_{old}$ are $\mathcal{N}_{i-new}$ and $\mathcal{N}_{i-old}$, respectively.

In each iteration, we first process the bidirectional insertion (Algorithm~\ref{HELP-Iteration}, \textit{Lines 11-14}) between nodes in $\mathcal{N}_{i-new}$ and $(\mathcal{N}_{i-new}\cup\mathcal{N}_{i-old})$, which are obtained from the previous iteration. In other words, the nodes in the aforementioned two sets are inserted into each other's neighbor sets, while maintaining that each node has at most $\varGamma$ neighbors. The aforementioned operation is illustrated in Fig.~\ref{fig:HELP-Index} (c). We stipulate that the maximum size of $\mathcal{N}_{i-new}$ is $\varGamma_{new}$, and $(\mathcal{N}_{i-new}\cup\mathcal{N}_{i-old})$ obtains up to $(\varGamma_{new}+\varGamma)$ nodes, then the complexity of \textit{Lines 7-14} is $O(N\varGamma_{new}(\varGamma_{new}+\varGamma))$.

After that, as shown in Fig.~\ref{fig:HELP-Index} (a, b), for each node $D_i\in D$, we separately update the neighbor set ($\mathcal{N}_{i-new}$, $\mathcal{N}_{i-rnew}$) of $D_i$ in ($\mathcal{I}_{new}$, $\mathcal{I}_{rnew}$) according to each new-added neighbor nodes and new reverse neighbor nodes in ${\mathcal{I}}$. 
Meanwhile, $\mathcal{I}_{old}$ and $\mathcal{I}_{rold}$ are updated following the same paradigm as $\mathcal{I}_{new}$ and $\mathcal{I}_{rnew}$ (\textit{Lines 16-23}). Afterwards, union $\mathcal{I}_{new}$ with $\mathcal{I}_{rnew}$ and $\mathcal{I}_{old}$ with $\mathcal{I}_{rold}$ for the next round of iteration (\textit{Line 24}). 
\textcolor{r1}{
Finally, to monitor the convergence of the index construction, we define Graph Quality~\cite{graphQuality,nhq} ($\psi$) as the approximation precision of the constructed graph with respect to the ground truth $k$-nearest neighbor graph. Formally, for a sampled node set $S$:}
\begin{equation}
\textcolor{r1}{\psi = \frac{1}{|S|} \sum_{u \in S} \frac{|\mathcal{N}(u) \cap \mathcal{N}_{gt}(u)|}{K}}
\end{equation}
\textcolor{r1}{where $\mathcal{N}(u)$ is the neighbor set in the current index and $\mathcal{N}_{gt}(u)$ is the ground truth $K$-nearest neighbors. And at the end of each iteration, we re-estimate and update the value of $\psi$. Following~\cite{nhq,graphQuality}, we set the measurement criteria $\varPsi$ of Graph Quality to $0.8$ for efficiency-accuracy trade-off. 
}



\textbf{\textit{(3) Complete the iteration process.}} When $\psi$ reaches the threshold $\varPsi$, the iteration process is completed. Assuming that the maximum number of the iteration rounds is \u{l}, the complexity of the complete iteration is $O($\u{l}$N \varGamma_{new}(\varGamma_{new}+\varGamma))$. 


\textbf{\textit{(4) Heterogeneous semantic prune.}} Inspired by~\cite{SSG}, we apply heterogeneous semantic pruning to eliminate redundant connections between nodes with similar semantics using a cosine\mbox{-}similarity threshold, while preserving connections between nodes with various attributes, as described in Algorithm~\ref{Heterogeneous-Semantic-Prune}. Through this operation, we can reduce the HELP index size to a certain extent. Thus, removing redundant connections reduces the number of nodes evaluated during subsequent routing, enhancing the search efficiency.

\setlength{\textfloatsep}{1pt}
\begin{algorithm}[ht] 
\small
    \caption{HELP Index Construction}
    \label{HELP-Iteration}
    \LinesNumbered
    \KwIn{Data Node Set $\mathcal{D}$, Maximum number of neighbors $\varGamma$, Maximum new-added neighbor number $\varGamma_{new}$, Cosine value threshold $\sigma$, and Graph quality threshold $\varPsi$.}

    \KwOut{HELP Index $\mathcal{I}=(D,E)$.}
    
    \textbf{Initialize:} Set an index $\mathcal{I}\!\!=\!\!(D,E)$, $D\!\!=\!\!\mathcal{D},E\!\!=\!\!\varnothing$.
    
    \For{each ${D_i}$ in $D$ }
    {
        Randomly generate set $\mathcal{N}_i$ of $\varGamma$ data nodes\;
        
        \lFor{each node ${n}$ in $\mathcal{N}_i$ }
        {
            $E = E\cup \overrightarrow{D_i,n}$;
        }
    }
   ${\mathcal{I}_{new}}$=$\mathcal{I}$; 
    $E_{old}, E_{rnew}, E_{rold}$=$\varnothing$;
    ${\mathcal{I}_{old}}$ =  $(D,E_{old})$, ${\mathcal{I}_{rnew}}$ =  $(D,E_{rnew})$, ${\mathcal{I}_{rold}}$ =  $(D,E_{rold})$;
    $\psi$=0;

    \While{$\psi<\varPsi$}{
        \textbf{parallel} \For{each ${D_i}$ in $\mathcal{D}$ }
        {
            $\mathcal{N}_{i-new}$, $\mathcal{N}_{i-old}$=$D_i$'s neighbor set in $\mathcal{I}_{new}$ and $\mathcal{I}_{old}$, respectively\;
            
            \For{each neighbor node $D_j$ in $\mathcal{N}_{i-new}$}{
                \For{each neighbor node $D_k$ in $(\mathcal{N}_{i-new} \cup \mathcal{N}_{i-old}-D_j)$}
                {
                    $D_x$ = $D_j$'s neighbor with the maximum distance;
                    $D_y$ = $D_k$'s neighbor with the maximum distance;
                    
                    $L_{jk}=\mathcal{U}(D_j,D_k)$;$L_{jx}=\mathcal{U}(D_j,D_x)$;$L_{ky}=\mathcal{U}(D_k,D_y)$\;
                    
                    \lIf{$L_{jk} < L_{jx}$}{
                        $E = E\cup \overrightarrow{D_jD_k}$; $E = E$\textbackslash $\overrightarrow{D_jD_x}$
                    }
                    \lIf{$L_{jk} < L_{ky}$}{
                        $E = E\cup \overrightarrow{D_kD_j}$; $E = E$\textbackslash $\overrightarrow{D_kD_y}$
                    }
                }
            }
        }
        $E_{new}, E_{old}, E_{rnew}, E_{rold}$ = $\varnothing$\;
        
        \textbf{parallel} \For{each ${D_i}$ in $\mathcal{D}$ }
        {
            $\mathcal{N}_i$ = $D_i$'s neighbor set in $\mathcal{I}$; $\gamma$ = 0\;
            
            \For{each neighbor node $D_j$ in $\mathcal{N}_i$}
            {

                \uIf{$D_j$ is a new-added node}{
                    $E_{new} = E_{new}\cup \overrightarrow{D_iD_j}$; 
                    $E_{rnew} = E_{rnew}\cup \overrightarrow{D_jD_i}$; 
                    mark $D_j$ as an old-added node\;
                    
                    $\gamma$++\;
                    
                    \lIf{$\gamma\geq\varGamma_{new}$}
                    {
                        break
                    }
                }\lElse{        
                    $E_{old} = E_{old}\cup \overrightarrow{D_iD_j}$; 
                    $E_{rold} = E_{rold}\cup \overrightarrow{D_jD_i}$
                }
            }
        }
        $\mathcal{I}_{new}$ = $\mathcal{I}_{new} \cup \mathcal{I}_{rnew}$; $\mathcal{I}_{old}$ = $\mathcal{I}_{old} \cup \mathcal{I}_{rold}$;
        Re-estimating $\varPsi$ and updating $\psi$\;
    }
    \textit{Heterogeneous-Semantic-Prune}$(\mathcal{I}, \mathcal{D}, \sigma, \varGamma)$\;
    
    return $\mathcal{I}$;
\end{algorithm}
    

Specifically, for each node $D_i\in D$, we apply the prune function on its neighbor sets $\mathcal{N}_i$ (\textit{Lines 12-13} in Algorithm~\ref{Heterogeneous-Semantic-Prune}). In the prune function, we define the $Select$ set to obtain the final selected neighbor nodes. And for each node $D_p\in \mathcal{N}_i$ (the input neighbor set), $D_p$ can be added to $Select$ if and only if its cosine value with all the nodes that share the same attribute in $Select$ exceeds the threshold value $\sigma$ (\textit{Lines 1-11} in Algorithm~\ref{Heterogeneous-Semantic-Prune}). Then $\mathcal{N}_i$ is updated via $Select$. 


It should be noted that during the pruning process, we set up in-degree checks to ensure that certain nodes are not pruned, which is described in \textit{Line 6} of Algorithm~\ref{Heterogeneous-Semantic-Prune}. Assume that $D_k$ is the neighbor of $D_s$ and it has an in-degree of 1. If $\overrightarrow{D_sD_k}$ is pruned, then $D_k$ will become an isolated island and we cannot find it by routing, which destroys the connectivity of the graph. In view of this, it is valuable to implement the in-degree checking mechanism, which is ignored by most previous studies


Furthermore, to enhance the connectivity between two unidirectional connected nodes, for each neighbor $D_j$ of $D_i$, we try to add $D_i$ to $D_j$'s neighbor sets $\mathcal{N}_j$ and once the size of $\mathcal{N}_j$ reaches $\varGamma$, $\mathcal{N}_j$ is updated by processing the prune function (\textit{Lines 14-19} in Algorithm~\ref{Heterogeneous-Semantic-Prune}). The complexity of the pruning process is $O(N\varGamma^2(1+\varGamma))$ for the worst case that all neighbor nodes are selected to the $Select$. Once all of the above steps have been performed, the construction is complete.

\noindent \textcolor{r1}{\textit{\textbf{Discussion: Differentiation from Standard Paradigms.}}}
\textcolor{r2}{To ensure robust connectivity in hybrid spaces, our Heterogeneous Semantic Pruning (HSP) is explicitly designed to construct ``topological highways'' across disparate attribute clusters. By selectively retaining ``bridge nodes''---neighbors with mismatched attributes but high feature similarity---and enforcing in-degree safeguards (Line 6 in Algorithm~\ref{Heterogeneous-Semantic-Prune}), HSP prevents the graph from fragmenting into isolated sub-regions.}
\textcolor{r3}{This capability is particularly vital for handling non-uniform attribute distributions. In such skewed scenarios, nodes with minority attributes are easily isolated by the majority class, forming ``Attribute Islands''. In contrast, HSP preserves these critical bridges, ensuring navigability even across highly imbalanced attribute sub-spaces.}
\textcolor{r1}{This structural optimization fundamentally differentiates HELP from standard neighbor refinement frameworks like NN-Descent~\cite{kgraph} and HNSW~\cite{HNSW2020}. These methods rely on greedy geometric pruning, which inadvertently saturates neighbor lists with nodes sharing identical attributes (due to their naturally smaller collaborative distances). Consequently, they sever the cross-attribute links essential for hybrid routing, leading to the formation of ``Attribute Islands'' where search algorithms easily become trapped.} \textcolor{r3}{Furthermore, regarding memory scalability, HELP maintains a flat graph structure with a space complexity of $O(N \cdot \Gamma)$. This is significantly more efficient than hierarchical designs like ACORN~\cite{acorn} (based on HNSW), which incur $O(N \cdot \Gamma \cdot (average~Layers))$ complexity to store multi-layer structures. Compared to methods based on NN-Descent, such as NHQ~\cite{nhq}, which also has time complexity $O(N \cdot \Gamma)$, we maximize the semantic utility of every stored connection by rigorously removing redundant edges via HSP, ensuring superior connectivity without the memory bloat.}

\begin{algorithm}[h]
\small
		\caption{Heterogeneous-Semantic-Prune}
		\label{Heterogeneous-Semantic-Prune}
		\LinesNumbered
		\KwIn{HELP Index $\mathcal{I}=(D,E)$, data Node Set $\mathcal{D}$, The cosine value threshold $\sigma$, Maximum number of neighbors $\varGamma$.}
		

    \SetKwProg{Fn}{Function}{:}{end}
    \Fn{Prune($\mathcal{N}$, $\sigma$, $\varGamma$, $D_s$)}{
		$Select$=$[\mathcal{N}[0]$]\;
        
		\For{each neighbor node $D_p$ in $\mathcal{N}$}
		{
			IfPruned = False\;
            
			\For{each node $D_k$ in $Select$}{
                \lIf{\textit{In-degree}$(D_k) = 1$}{
					break //Guarantee that the in-degree of $D_k$ is at least 1, i.e., it's a neighbor of at least one node
				}
				\lIf{$\sigma < cos$$<\overrightarrow{D_sD_p},\overrightarrow{D_sD_k}>$ $ \&\& A_p=A_k $}{
					IfPruned=True
				}
			}
			\uIf{$\rm !IfPruned$}{

				\lIf{$Select$.size  $<\varGamma$}{
					$Select$.add($D_p$)
				}
                    \lIf{$Select$.size  $=\varGamma$}{
                        break
                    }
			}
		}
		return $Select$;
    }

		\textbf{parallel} \For{each ${D_i}$ in $\mathcal{D}$}
		{
			$\mathcal{N}_i$=$D_i$'s neighbor set in $\mathcal{I}$;
			$\mathcal{N}_i$=\textit{Prune}$(\mathcal{N}_i, \sigma, \varGamma, D_i)$\;
		}
		
		\textbf{parallel} \For{each ${D_i}$ in $\mathcal{D}$}
		{
			$\mathcal{N}_i$ = $D_i$'s neighbor set in $\mathcal{I}$\;
            
			\For{each neighbor node $D_j$ in $\mathcal{N}_i$}
			{
				$\mathcal{N}_j$ = $D_j$'s neighbor set in $\mathcal{I}$;
				$\mathcal{N}_j$ = $\mathcal{N}_j \cup D_i$\;
                
				\uIf{$\mathcal{N}_j$.size  $>\varGamma$}{
					$\mathcal{N}_j$ = \textit{Prune}$(\mathcal{N}_j, \sigma, \varGamma, D_j)$\;
				}
				
			}
		}
\end{algorithm}

\subsection{Dynamic Heterogeneity Routing}
\label{Local Routing Process}

\textcolor{r4}{To maximize retrieval efficiency, our routing strategy employs a ``Coarse-to-Fine'' acceleration mechanism.~(1)~Dynamic Coarse Routing~(Rapid Approach): This phase serves as a lightweight accelerator by maintaining a compact pioneer set. By updating this small-sized set via partial neighbor inspections, the algorithm rapidly traverses the graph to the approximate vicinity of the query with minimal computational overhead. (2)~Greedy Refinement Routing (Precise Convergence): Once the pioneer set stops updating, it indicates the candidate set is anchored near the target. The strategy switches to a fine-grained mode, employing a greedy approach to exhaustively evaluate all neighbors of candidate nodes to precisely identify the final Top-K nodes. This decomposition significantly reduces the total number of distance calculations, ensuring high throughput without sacrificing accuracy.}

The overall procedure is summarized in Algorithm~\ref{One-Step}, which comprises the following three main operations.
\textbf{\textit{(1) Initialization.}} In this operation, we define two node sets: result set~($\mathcal{R}$) and pioneer set~($\mathcal{P}$). The former contains the desired top-$K$ nodes for the given query $\mathcal{Q}$, and the latter contains the seed nodes for routing. We randomly select $K$ nodes from the HELP index, and sort these nodes in distance ascending order to $\mathcal{Q}$, thereby forming $\mathcal{R}$. Thereafter, we select the first $P$ nodes from $\mathcal{R}$ to generate $\mathcal{P}$. 
	
\textbf{\textit{(2) Dynamic Coarse Routing.}} For each $D_a$ in $\mathcal{P}$, we choose \textbf{\textit{half}} the number of its neighbors for the coarse routing (\textit{Lines 2-10} of Algorithm~\ref{One-Step}) to rapidly identify the seed nodes. Specifically, we sequentially evaluate the similarity between the neighbors of the node $D_a$ in $\mathcal{P}$ ($D_b$, $D_c$, etc.) and query $\mathcal{Q}$. If $D_b$ is more similar to $\mathcal{Q}$ than the least similar node in $\mathcal{P}$, we employ the insertion sorting algorithm to insert it into $\mathcal{R}$ and $\mathcal{P}$. 
 The above process continues until no more new nodes can be inserted into $\mathcal{P}$.

\setlength{\textfloatsep}{0pt}

\begin{algorithm}[ht]
    \small
	\caption{Dynamic Heterogeneity Routing}
	\label{One-Step}
	
	\LinesNumbered

    \KwIn{Target number $K$, pioneer set size $P$, Query $\mathcal{Q}$, Index $\mathcal{I}$, \textcolor{r1}{and the AUTO metric $\mathcal{U}(\cdot,\cdot)$ defined in Eq.~\ref{fd}.}}
	
	\KwOut{$\mathcal{R}$ which contains the top-$K$ nodes for $\mathcal{Q}$.}
	
	\textbf{Initialize:}Randomly select $K$ nodes from $\mathcal{D}$ and sort these nodes in ascending order of distance to $\mathcal{Q}$ to form $\mathcal{R}=\{D_k\}_{k=1}^{K}$; Choose the first $P$ nodes from $\mathcal{R}$ to generate $\mathcal{P}=\{D_p\}_{p=1}^{P}$. $j$ = 0 \;

	\While{unchecked $D_p$ exists in $\mathcal{P}$}{
		
		mark $D_p$ as checked;$s$=$D_p$'s half neighbor number\;
		
		\For{$j<$ s}
		{
			
			$N_j$ = Search($\mathcal{I}$,~$D_p$,~$j$); // find the $j$-th neighbor node of $D_p$ within the index $\mathcal{I}$
			
				
			
				

            \uIf{\textcolor{r1}{$\mathcal{U}(N_j, Q) < \mathcal{U}(D_K, Q)$}}{
				
				Orderly insert $N_j$ into $\mathcal{R}$;\
			}
			
			\uIf {\textcolor{r1}{$\mathcal{U}(N_j, Q) < \mathcal{U}(D_P, Q)$}}{
				
				Orderly insert $N_j$ into $\mathcal{P}$;\ // $D_K$ and $D_P$ are the last node in $\mathcal{R}$ and $\mathcal{P}$, respectively.
			}
			
			$j = j + 1$\;
			
		}
		
	}
	$j$ = 0\;
    
	\While{ unchecked $D_k$ exists in $R$}{
		
		mark $D_k$ as checked;
		$s$=neighbor number of $D_k$\;
		
		\For{$j<$ s}
		{
			
			$N_j$ = Search($\mathcal{I}$,~$D_k$,~j);
			

            \uIf{\textcolor{r1}{$\mathcal{U}(N_j, Q) < \mathcal{U}(D_K, Q)$}}{
				Orderly insert $N_j$ into $\mathcal{R}$;\
			}
			
			$j = j + 1$\;
			
		}
		
	}
	
	return $\mathcal{R}$;
	
    \end{algorithm}

\textbf{\textit{(3) Greedy Refinement Routing.}} After having the seed nodes in $\mathcal{R}$, we perform the greedy refinement routing on all the nodes in $\mathcal{R}$ (\textit{Lines 11-18} of Algorithm~\ref{One-Step}). Concretely, for each node $D_i$ in $\mathcal{R}$, we iteratively evaluate the similarity of its \textbf{\textit{all}} neighbors with $\mathcal{Q}$. If a neighbor is more similar to $\mathcal{Q}$ than $D_i$, it is inserted into $\mathcal{R}$ using the insertion sorting algorithm. 
 After all nodes in $\mathcal{R}$ have been validated, the remained nodes in $\mathcal{R}$ are considered to be the top-$K$ results for the given query $\mathcal{Q}$, thereby completing hybrid ANNS. Assuming that the path length of the dynamic coarse routing is $l_\mathcal{P}$, the time complexity of our greedy refinement routing is $O(\varGamma(log N - l_\mathcal{P}))$, outperforming the $O(\varGamma logN)$ complexity of the mainstream baseline methods~\cite{NSG,HNSW2020}. Subsequent comparative experiments will illustrate the superior retrieval efficiency of our approach.

\noindent \textit{\textbf{\textcolor{r2}{Discussion: Advantage over Existing Routing Strategies.}} }
\textcolor{r2}{To ensure efficient navigation, our Dynamic Heterogeneity Routing adopts a ``Coarse-to-Fine'' strategy. By innovatively maintaining a dynamic Pioneer Set (Algorithm~\ref{One-Step}), the algorithm prioritizes a coarse-grained global probing phase. This allows it to rapidly locate the target neighborhood with minimal computational overhead before switching to fine-grained refinement.}
\textcolor{r3}{Theoretically, our Dynamic Heterogeneity Routing optimizes the search complexity to \textbf{$O(\Gamma(\log N - l_{\mathcal{P}}))$}, where $l_{\mathcal{P}}$ denotes the path length skipped by the pioneer set. This formulation underscores our scalability advantage:}
\textcolor{r3}{\textbf{(1) vs. NHQ}~\cite{nhq}: While NHQ shares the complexity of $O(\Gamma(\log N - l_{\mathcal{P}}))$, its strict first-stage exit condition constrains the timing of granularity adjustments, leading to redundant computations in subsequent stages. In contrast, our dynamic strategy significantly boosts QPS, as shown in the sub-experiment \textbf{w/o~Dynamic} of the ablation study (Section~\ref{Ablation Analysis (RQ3)}, Fig.~\ref{fig:stress_test}).}
\textcolor{r3}{\textbf{(2) vs. ACORN}~\cite{acorn}: This predicate-agnostic method theoretically operate in $O(\Gamma_{eff} \cdot \log N)$. However, under high attribute cardinality, the valid subgraph becomes sparse, causing the effective neighbor count $\Gamma_{eff}$ to inflate due to backtracking, resulting in significant efficiency degradation (as shown in Section~\ref{Retrieval Performance Comparison (RQ1)}, Fig.~\ref{figall}).}

\subsection{\textcolor{r1}{Extension to Subset Attribute Queries}}
\label{sec:subset_extension}

\textcolor{r3}{In real-world industrial scenarios, systems often face challenges such as data with missing values and user queries involving only a subset of attributes. STABLE inherently supports these scenarios through a Masking Mechanism without requiring modifications to the index structure.}
\textcolor{r1}{Specifically, we define a mask vector $\mathbf{m} \in \{0, 1\}^L$. For a query, $m_l=0$ indicates a wildcard (subset query); similarly, for data nodes, $m_l=0$ can denote a missing value in the $l$-th dimension. The attribute consistency calculation is adapted as:}
\begin{equation}
\textcolor{r1}{\mathcal{S}_{A}(A_{i},\widehat{A})=\sum_{l=1}^{L} m_l \cdot |a_{l}^{i}-\widehat{a}_{l}|}
\end{equation}
\textcolor{r3}{By incorporating this mask, STABLE dynamically ignores unspecified or missing attributes during evaluation. This allows a single unified index to support flexible queries and incomplete data, demonstrating superior practical adaptability compared to partition-based methods.}




\section{Experiments and evaluation}
\label{sec4}
In this section, we first outline the experimental settings. We then conduct key comparative experiments to evaluate the effectiveness of our method by addressing the following research questions (RQs).

\noindent
\textbf{$\bullet$~RQ1:}~Is retrieval performance~(Recall@$10$,$\!$ QPS) of STABLE superior to state-of-the-art competitors?


\noindent
\textbf{$\bullet$~RQ2:}~\textcolor{r2}{Is our STABLE robust to variations in data volume, attribute cardinality, and query selectivity?}

\noindent
\textbf{$\bullet$~RQ3:}~Is each major module of STABLE effective?

\noindent
\textbf{$\bullet$~RQ4:}~Is our index construction effective?

\noindent
\textbf{$\bullet$~RQ5:}~Is the computation of parameter $\alpha$ in AUTO metric validated to be effective for improving retrieval performance?

\noindent
\textbf{$\bullet$~RQ6:}~\textcolor{r1}{How do the key hyperparameters impact the retrieval performance and index size?}

\noindent
\textbf{$\bullet$~RQ7:}~\textcolor{r1}{Does our AUTO metric hinder the SIMD acceleration?}



\subsection{Experimental settings}

\underline{\textbf{Datasets:}} We select three categories of \textit{1M}-scale benchmark feature vector datasets, i.e., SIFT1M\footnote{http://corpus-texmex.irisa.fr/}, GLOVE-100\footnote{https://nlp.stanford.edu/projects/glove/} and CRAWL\footnote{\label{fn:crawl}https://commoncrawl.org/}. Considering the proliferation of data volumes recently, to test whether our method can migrate to larger datasets, we also leverage two \textit{10M}-scale benchmark feature vector datasets for further experiments, i.e., BigANN10M\footnote{https://big-ann-benchmarks.com/neurips21.html} and DEEP10M\footnote{https://research.yandex.com/blog/benchmarks-for-billion-scale-similarity-search}~\cite{DEEP1B}. 
Following previous works~\cite{nhq,filterDiskann}, we employ a simple attribute generation strategy to construct attribute constraints with three levels of attribute cardinality for each dataset category, thereby obtaining nine \textit{1M}-scale datasets and six \textit{10M}-scale datasets for our experiments. More details of these datasets are listed in TABLE~\ref{datasets}.

\begin{table}[h]
    \centering
    \vspace{-10pt}
    \caption{The specific statistics of datasets. \textit{The symbols $M$ and $L$ are the dimension numbers of feature vectors and attribute vectors. $N$ and $\Theta$ denote the node number and the attribute cardinality, respectively.}}
    \vspace{-6pt}
    \tabcolsep=4pt
	\resizebox{\linewidth}{!}{
		\begin{tabular}{c|c|c|c|c|c|c} 
		\Xhline{1pt}
		\hline\hline
			\textbf{Datasets}&Category&\textbf{$M$}& \textbf{$L$}& \textbf{$\Theta$}&	\textbf{$N$}& \textbf{Query Num} 	\\ \hline	
			\rowcolor[rgb]{ .949,  .949,  .949} \multicolumn{7}{c}{\textit{1M-Scale Datasets}} \\
			SIFT-7-3&\multirowcell{3}{SIFT1M} 
			& 	&7& 	2187& 	&    \\ \cline{1-1} \cline{4-5}
			SIFT-6-3&& 	128&6&	729& 	1,000,000& 	10,000   \\ \cline{1-1}\cline{4-5}
			SIFT-5-3&& 	&5& 	243& 	&   \\ \hline 
			
			GLOVE-7-3&\multirowcell{3}{GLOVE-100} 
			&  &7	& 	2187& 	&    \\ \cline{1-1} \cline{4-5}
			GLOVE-6-3&& 	100&6&	729& 	1,183,514& 	10,000   \\ \cline{1-1}\cline{4-5}
			GLOVE-5-3&& 	&5& 	243 & 	&   \\ \hline 
			
			CRAWL-7-3&\multirowcell{3}{CRAWL} 
			& 	&7& 	2187 & 	&    \\ \cline{1-1} \cline{4-5}
			CRAWL-6-3&& 	300&6&	729& 	1,989,995& 	10,000   \\ \cline{1-1}\cline{4-5}
			CRAWL-5-3&& 	&5& 	243& 	&   \\ 		
			\rowcolor[rgb]{ .949,  .949,  .949} \multicolumn{7}{c}{\textit{10M-Scale Datasets}} \\
            BigANN-7-3&\multirowcell{3}{BigANN10M} 
			&  &7	& 	2187& 	&    \\ \cline{1-1} \cline{4-5}
			BigANN-6-3&& 	128&6&	729& 	10,000,000& 	10,000   \\ \cline{1-1}\cline{4-5}
			BigANN-5-3&& 	&5& 	243 & 	&   \\ \hline 
            DEEP-7-3&\multirowcell{3}{DEPP10M} 
			&  &7	& 	2187& 	&    \\ \cline{1-1} \cline{4-5}
			DEEP-6-3&& 	96&6&	729& 	10,000,000& 	10,000   \\ \cline{1-1}\cline{4-5}
			DEEP-5-3&& 	&5& 	243 & 	&   \\
            
		\hline\hline
			\Xhline{1pt}
		\end{tabular}
	}
 \label{datasets}
 \end{table}

\textit{Attribute Generation Strategy and Attribute Cardinality $\Theta$.} Furthermore, to better represent the intensity of the attribute constraints we add, we numerically measure them using the attribute cardinality, dubbed as $\Theta$. 
We take CRAWL-5-3 as an example to explain our attribute generation strategy and describe how to obtain the attribute cardinality. First, we collect data from the \textit{Common Crawl}\footref{fn:crawl} to form our label pool. Subsequently, for CRAWL-5-3, we generate a $5$-dimensional attribute vector~(e.g., ``content publishing time'', ``page authority'', ``content types'', ``character encodings'' and ``languages'') corresponding to each feature vector in CRAWL, where each dimensional attribute value is assigned a label from the label pool of size $3$~(e.g., ``page authority'' owns $3$ values: ``authority sites'', ``spam sites'', ``user-generated content''). And then the $\Theta$ of CRAWL-5-3 is computed by $3^5=243$.
With the continuous growth of $\Theta$, conducting hybrid ANNS involves processing more complex attribute semantic information.
\textcolor{r2}{Correspondingly, to comprehensively evaluate retrieval performance, we configure the query settings with varying degrees of complexity. Specifically, for our main comparative experiments, we focus on rigorous conditions where queries involve 5 to 7 active attribute filters (denoted as $F$). Given the size of label pool ($3$), this setup enforces extremely low selectivity~(the proportion of nodes with perfectly matched attributes), calculated approximately as $(1/3)^F$~(Although actual selectivity fluctuates due to the natural skewness of crawled data, this metric effectively captures the exponential decay in result set size as filter complexity increases.). Consequently, the selectivity ranges from $\approx 0.41\%$ (calculated as $1/3^5$ for $F=5$) down to $\approx 0.046\%$ (calculated as $1/3^7$ for $F=7$). This configuration ensures that our main experiments prioritize evaluating the methods' robustness under severe Connectivity Barriers.
Furthermore, to verify the broad applicability of STABLE across different query scenarios, we conduct an additional Stress Test covering the full range of $F$ from 1 to 7, spanning from loose ($\approx 33.3\%$) to strict filtering conditions.}

\begin{figure*}[ht]
 \vspace{-10pt}
\begin{center}
    	\includegraphics[width=\linewidth]{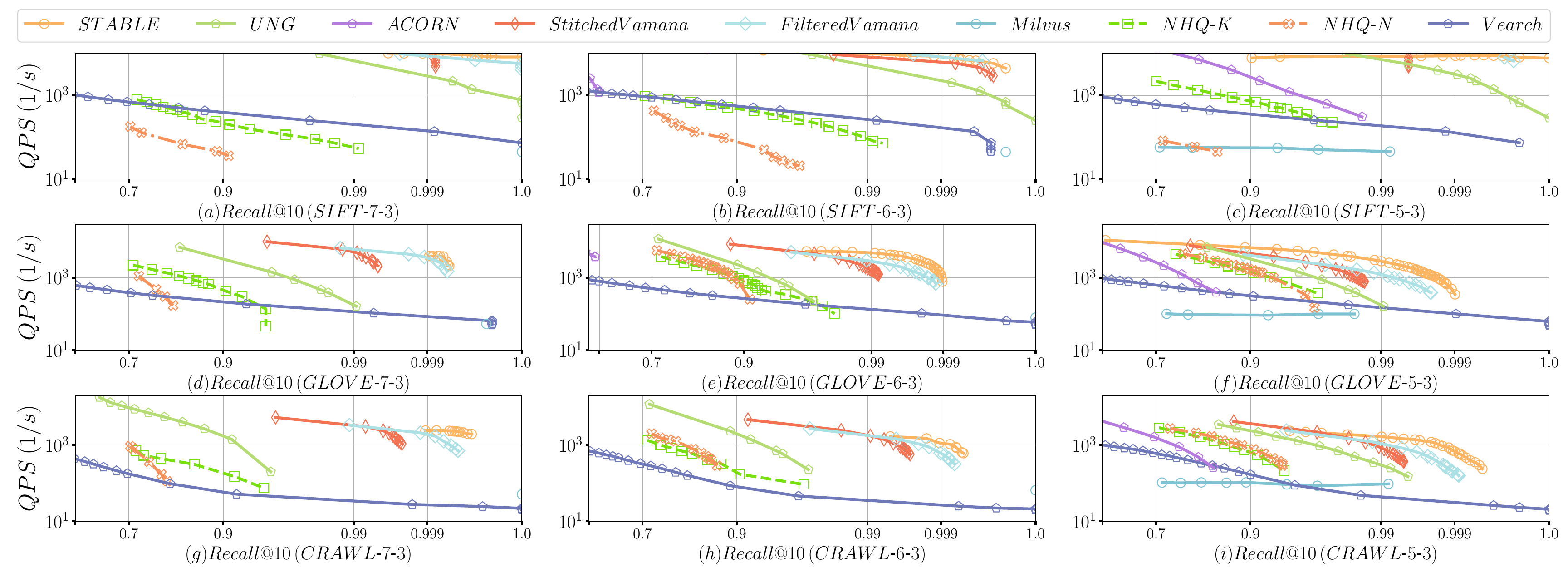}
\end{center}
 \vspace{-6pt}
	\caption{(a)-(i) represent the QPS vs Recall@$10$ performance of hybrid ANNS methods on nine \textit{1M}-scale datasets.}
	\label{figall}
\vspace{-18pt}
\end{figure*} 

\noindent
\underline{\textbf{Compared Methods:}} We select seven state-of-the-art hybrid ANNS algorithms as benchmarks to verify the effectiveness of our method.

\noindent
\textbf{$\bullet\,$Milvus}~\cite{milvus2021}: it is a \textit{Pre-filtering} hybrid ANNS method. We first implement Milvus~\cite{milvus2021} and then perform hybrid ANNS following its attribute filtering and feature similarity calculation guidelines.
	
\noindent
\textbf{$\bullet\,$Vearch}~\cite{vearch}: it is a \textit{Post-filtering} based hybrid ANNS method proposed by JD company. We first implement Vearch~\cite{vearch} and then follow its original similarity evaluation and attribute filtering rules for subsequent hybrid ANNS.

\noindent
\textbf{$\bullet$NHQ-K}~\cite{nhq}: it is an efficient attempt to apply NHQ~\cite{nhq} to the KGraph~\cite{kgraph} index, enabling the achievement of \textit{Specialized Indices} for Hybrid ANNS. We first implement NHQ-K and then follow its original experimental settings~\cite{nhq} for hybrid ANNS. 
    
\noindent
\textbf{$\bullet\,$NHQ-N}~\cite{nhq}: it is a \textit{Specialized Index} method for hybrid ANNS and can jointly consider feature similarity and attribute consistency. It constructs NSW~\cite{nsw} graph index for retrieval. We first implement NHQ-N and then follow its original experimental settings~\cite{nhq} for hybrid ANNS.

\noindent
\textbf{$\bullet\,$FilteredVamana}~\cite{filterDiskann}: it incrementally connects semantically similar nodes to construct the graph index, and subsequently completes hybrid ANNS via predefined index entry nodes. We implement it while maintaining the original parameter settings~\cite{filterDiskann}.


\noindent
\textbf{$\bullet\,$StitchedVamana}~\cite{filterDiskann}: it takes a bulk-indexing approach for the graph index construction and then conducts hybrid ANNS via predefined entry nodes. We follow its original parameter settings~\cite{filterDiskann} to implement StitchedVamana.
    
\noindent
\textbf{$\bullet\,$ACORN}~\cite{acorn}: It designs the \textit{Specialized Index} method for performant and predicate-agnostic hybrid search based on the HNSW index. We implement it via its official code~\cite{acorn}.

\noindent
\textcolor{r1}{
\textbf{$\bullet\,$UNG}~\cite{ung}: It is a \textit{Specialized Index} method via \textit{Unified Navigation}. We implement it via its official code~\cite{ung}.}

\noindent
\underline{\textbf{Metrics:}} Aiming for objective performance comparisons on the above datasets, we first select the \textit{Queries Per Second} (QPS) and Recall@$\mathcal{K}$ (abbreviated as R@$\mathcal{K}$) to assess the retrieval efficiency and accuracy. Concretely, Recall@$\mathcal{K}$ is calculated as $Recall@\mathcal{K} = \frac{|R_\mathcal{K} \cap R_T|}{\mathcal{K}}$,
where $R_\mathcal{K}$ denotes the retrieved top-$\mathcal{K}$ results and $R_T$ is the ground truth results. In accordance with widely used standards, we set $\mathcal{K}=10$.
Subsequently, we select the average build time (\textit{seconds}) to evaluate the index construction efficiency.

\noindent
\underline{\textbf{Implementation setup:}} All codes of STABLE are written in C++ and compiled via $CMake$ $3.18$. All experiments are performed on a server with Ubuntu $18.04$, an Intel(R) Xeon(R) Platinum 8275CL CPU at $3.00$GHz, and $256$G memory. We utilize $8$ threads for index construction and a single thread for the hybrid ANNS. The maximum neighbor number $\varGamma$ and the maximum new-added neighbor number $\varGamma_{new}$ are set to $100$ on SIFT1M, BigANN10M, and DEEP10M, and $120$ on GLOVE-100 and CRAWL. The cosine value threshold $\sigma$ is set to $0.44$ on SIFT1M, BigANN10M, and DEEP10M, and $0.49$ on GLOVE-100 and CRAWL. The graph quality threshold $\varPhi$ is set to $0.8$ on all datasets. The target number $K$ in Algorithm~\ref{One-Step} is set from $10$ to $500$ to obtain the \textit{QPS vs Recall} plots and the pioneer set size $P=K/2$.

\subsection{Retrieval Performance Comparison (RQ1)}
\label{Retrieval Performance Comparison (RQ1)}
In this subsection, we compare the retrieval performance of STABLE against seven baseline methods on \textit{1M}-scale datasets, evaluating both accuracy and efficiency. We generate \textit{QPS vs Recall} plots based on the generic evaluation strategy~\cite{nhq,filterDiskann,acorn}. In these plots, each curve corresponds to a hybrid ANNS method, and each point indicates the Recall@$10$ (x-axis) and the corresponding QPS efficiency (y-axis). According to the results in Fig.~\ref{figall} on the nine \textit{1M}-scale datasets, we obtain the following observations.
\textbf{(1)} Milvus performs worst on the $9$ datasets, likely due to its partitioned retrieval strategy. The attribute filtering with larger $\Theta$ leads to many attribute partitions with few feature vectors in each, prompting Milvus to use brute search within each partition, which improves recall but reduces efficiency. As shown in Fig.~\ref{figall}, except for SIFT-5-3 (Fig.~\ref{figall}(c)), GLOVE-5-3 (Fig.~\ref{figall}(f)), and CRAWL-5-3 (Fig.~\ref{figall}(i)), Milvus uses brute search on the $6$ datasets, achieving high Recall@$10$ with low QPS. Smaller $\Theta$ results in fewer partitions with more vectors, allowing Milvus to index partitions and use its routing strategy, but its performance remains unsatisfactory compared to state-of-the-art methods, as shown in Fig.~\ref{figall}~(c, f, i). 
\textcolor{r2}{Notably, in these cases (e.g., $\Theta=243$), the Milvus performance curve manifests as a quasi-straight line. This phenomenon occurs because the data is fragmented into small partitions (approx. 4,000 vectors), causing the search latency to be dominated by fixed system overheads (e.g., partition location and attribute filtering) rather than vector computation. Consequently, the retrieval speed becomes insensitive to index parameter tuning, limiting the trade-off space between efficiency and accuracy.}
\textbf{(2)} Vearch outperforms Milvus in retrieval performance, but the improvement is limited due to the separate execution of feature similarity calculation and attribute filtering, as shown in Fig.~\ref{figall}. Specifically, to achieve high accuracy, Vearch relaxes the similarity evaluation criteria to select more candidates for attribute filtering, but increasing the number of candidates negatively impacts retrieval efficiency.
\textbf{(3)} NHQ-K and NHQ-N achieve superior retrieval performance than above two baselines, demonstrating that their collaborative semantic measurement and routing strategy enhance hybrid ANNS. However, as attribute cardinality increases (with larger $\Theta$), their performance significantly declines. As shown in Fig.~\ref{figall}~(i, h, g), both NHQ-K and NHQ-N experience substantial reductions in Recall@$10$ and QPS, indicating that they lack robustness to changes in attribute cardinality.
\textbf{(4)} FilteredVamana and StitchedVamana achieve higher Recall@$10$ and QPS than the other $5$ baselines, indicating that their semantic connection index and entry nodes-based routing enhance retrieval performance. However, as shown in Fig.~\ref{figall}(c), their performance bottlenecks occur earlier than STABLE, suggesting that as attribute cardinality increases (with larger $\Theta$), their semantic comprehension abilities degrade more quickly than STABLE, limiting both retrieval accuracy and efficiency.
\textbf{(5)} Similarly, ACORN struggles with variations in attribute cardinality. Specifically, it achieves high recall only on the dataset with $\Theta=243$, while its recall remains below $0.6$ on other datasets, as shown in Fig.~\ref{figall}(b). This may be due to ACORN's predicate-subgraph traversal being sensitive to attribute cardinality changes, making it challenging to maintain robust retrieval performance.
\textcolor{r1}{Unlike ACORN, UNG exhibits performance volatility when handling diverse feature distributions. As illustrated in Fig.~\ref{figall}, while UNG achieves superior performance on the SIFT1M dataset, it suffers a marked decline on GLOVE-100 and CRAWL compared to competing methods. The performance fluctuations observed in both ACORN and UNG corroborate our analysis in Challenge 1~(C1) that the lack of flexible design constitutes a critical bottleneck constraining robustness across diverse datasets.}
\textbf{(6)} STABLE outperforms all methods across all datasets and demonstrates robust retrieval performance under various attribute cardinalities. This verifies that our STABLE consisting of AUTO metric, HELP index, and dynamic heterogeneity routing is indeed practical and robust for hybrid ANNS, thereby noticeably improving the retrieval performance.

\subsection{Analysis on Robustness for Data Volume and Attribute Cardinality (RQ2)}

\textcolor{r2}{In this subsection, we conduct a comprehensive robustness analysis of STABLE from three perspectives: data volume scalability on \textit{10M}-scale datasets, adaptability to data distribution via various attribute cardinalities, and stability under different query selectivities.}



\begin{figure*}[ht]
 \vspace{-11pt}
\begin{center}
    	\includegraphics[width=\linewidth]{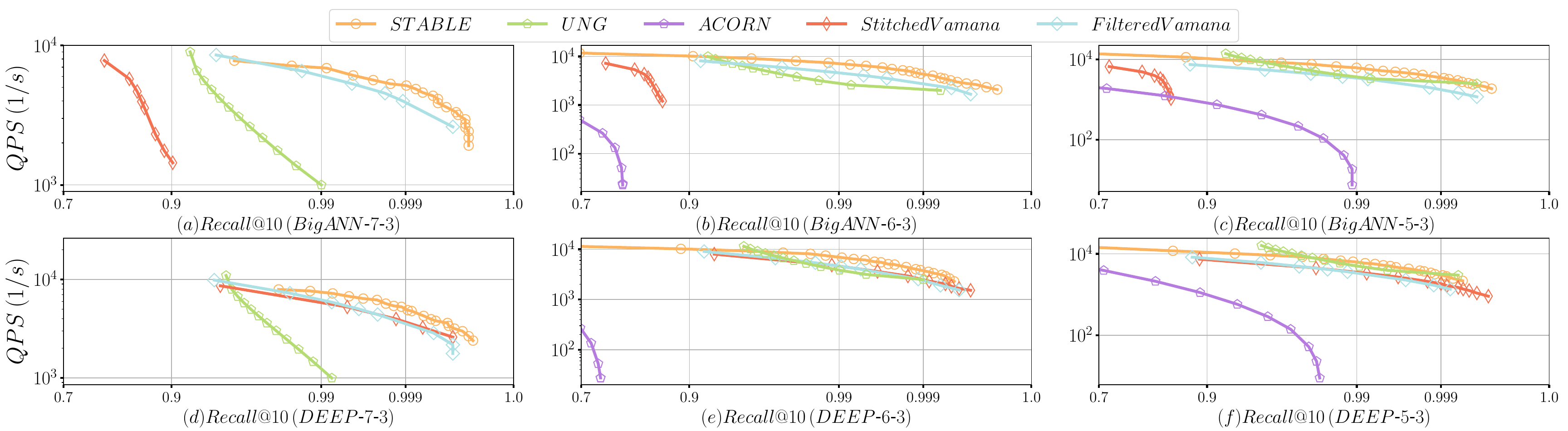}
\end{center}
 \vspace{-6pt}
	\caption{(a)-(f) represent the QPS vs Recall@$10$ performance of hybrid ANNS methods on six \textit{10M}-scale datasets.}
	\label{fig10M}
\vspace{-18pt}
\end{figure*} 

\underline{\textbf{\textit{On Robustness for Data Volume.}}}
To validate the robustness and scalability of STABLE on the datasets with more than the \textit{1M}-scale, we further compared the retrieval performance of STABLE on six \textit{10M}-scale datasets with representative baselines on the \textit{1M}-scale datasets, i.e., ACORN, UNG, FilteredVamana, and StitchedVamana. From the results illustrated in Fig.~\ref{fig10M}, we can derive the following observations.
\textbf{(1)} ACORN remains sensitive to attribute cardinality at the \textit{10M}-scale datasets. Although it yields slightly higher recall than at \textit{1M}-scale with cardinality $\Theta=243$, its subgraph traversal struggles when partitions contain few vectors. 
\textcolor{r1}{\textbf{(2)} UNG similarly exhibits sensitivity to attribute cardinality. This is likely because the Unified Navigation paradigm struggles to maintain robust graph connectivity when handling complex multi-attribute and multi-vector distributions, resulting in performance fluctuations.}
\textbf{(3)} StitchedVamana obtains the inferior performance on the BigANN10M dataset, and compared to the \textit{1M}-scale dataset, its retrieval accuracy reveals a significant decline. As depicted in Fig.~\ref{fig10M}(a-c), the retrieval recalls of StitchedVamana only reach around $0.9$. This may be due to that its bulk-indexing approach suffers from limitations in scalability and cannot handle the data distribution of BigANN10M well, which makes it difficult to achieve the desired accuracy. 
\textbf{(4)} FilteredVamana demonstrates fair scalability on large-scale datasets. However, its retrieval performance is still sub-optimal, which may be due to its limited heterogeneous semantic perception.
\textbf{(5)} Our STABLE continues to demonstrate optimal retrieval performance on the \textit{10M}-scale datasets, which substantiates its superior scalability and effectiveness.

\underline{\textbf{\textit{On Robustness for Attribute Cardinality.}}} In our robustness evaluation, we assess STABLE and three representative baselines, ACORN~\cite{acorn}, StitchedVamana~\cite{filterDiskann}, and FilteredVamana~\cite{filterDiskann} across six attribute cardinality $\Theta$. For simplicity, we reassign labels to obatin various attribute cardinality settings.
TABLE~\ref{robust} reports Recall@$10$ at a fixed QPS=$3000$, a value supported by current methods and serving as an approximate upper benchmark. Lower QPS settings provide limited insight, as they correspond to reduced retrieval efficiency. From these results, we draw the following observations.

\begin{table}[h]
  \centering
  \vspace{-6pt}
  \caption{We fix QPS values around $3000$ and report corresponding R@$10$ values on SIFT1M with various $\Theta$s. The overall best results are in bold.}
    \vspace{-4pt}
      \tabcolsep=2pt
  	\resizebox{\linewidth}{!}{
    \begin{tabular}{c|c|c|c|c|c|c}
    \Xhline{1pt}
    \hline\hline
           \rowcolor[rgb]{ .949,  .949,  .949}
    \textit{Fixed QPS$=3000$} & $\Theta$=$3000$ & $\Theta$=$2000$ & $\Theta$=$1000$ & $\Theta$=$500$ & $\Theta$=$100$ & $\Theta$=$50$ \\
    \hline
    Algorithms & R@$10$ & R@$10$ & R@$10$ & R@$10$ & R@$10$ & R@$10$ \\
    \hline
    StitchedVamana~\cite{filterDiskann}~\footnotesize{\textcolor{gray}{(WWW'23)}} & 0.99530 & 0.99890 & 0.99900 & 0.99995 & 0.99752 & 0.98909 \\
    FileredVamana~\cite{filterDiskann}~\footnotesize{\textcolor{gray}{(WWW'23)}} & \textbf{1.00000}     & \textbf{1.00000}     & \textbf{1.00000}     & 0.99993 & 0.99700 & 0.98224 \\
    ACORN~\cite{acorn}~\footnotesize{\textcolor{gray}{(SIGMOD'24)}} & 0.00747 & 0.02065 & 0.28122 & 0.70018 & 0.97537 & 0.98851 \\
    \hline
    \rowcolor[rgb]{ .851,  .851,  .851}
    \textbf{STABLE (Ours)} & \textbf{1.00000}     & \textbf{1.00000}     & \textbf{1.00000}     & \textbf{0.99997\textuparrow} & \textbf{0.9989\textuparrow} & \textbf{0.99702\textuparrow} \\
    \hline\hline
    \Xhline{1pt}

    \end{tabular}%
    }
	\label{robust}
    \vspace{-8pt}
\end{table}

\begin{figure*}[h]
 \vspace{-12pt}
 \begin{center}
     \includegraphics[width=\linewidth]{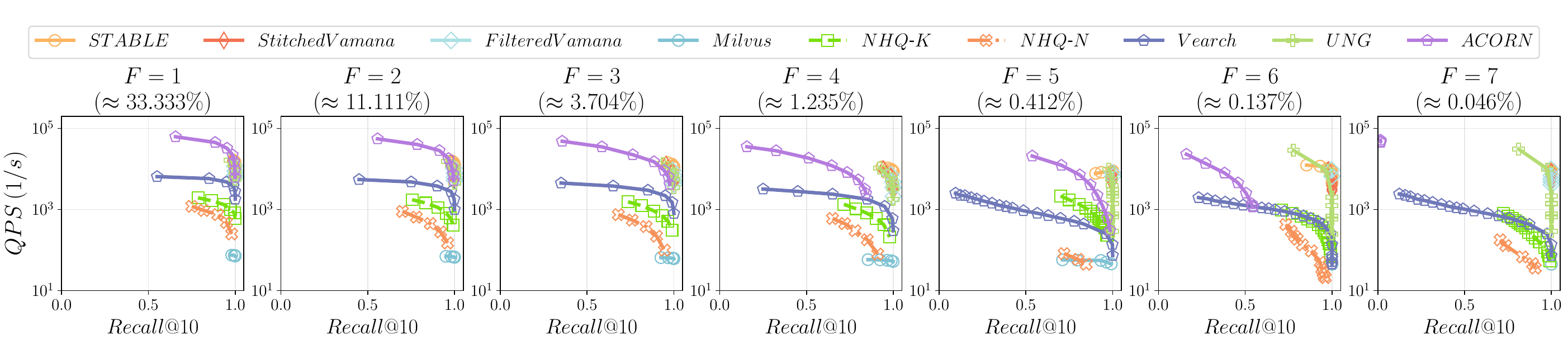}
 \end{center}
 \vspace{-6pt}
	\caption{Performance comparison under varying query selectivities on SIFT-7-3 dataset. The subplots display QPS vs. Recall@10 as the number of active filters ($F$) increases from 1 to 7. The percentage in parentheses indicates the approximate selectivity.}
	\label{fig:stress_test}
  \vspace{-10pt}
\end{figure*}

\begin{figure*}[h]
 \vspace{-2pt}
 \begin{center}
     \includegraphics[width=\linewidth]{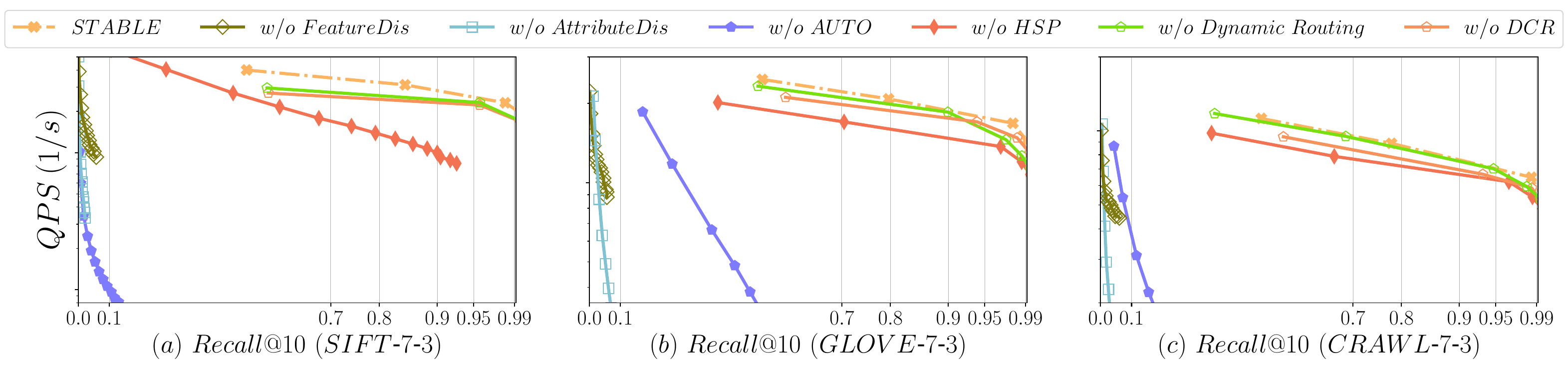}
 \end{center}
 \vspace{-6pt}
	\caption{The Ablation Studies of STABLE with different settings on (a) SIFT-7-3, (b) GLOVE-7-3, and (c) CRAWL-7-3 datasets.}
	\label{ablation}
  \vspace{-18pt}
\end{figure*}

At a fixed QPS of $3000$, STABLE achieves superior retrieval accuracy across all attribute cardinalities, with all Recall@$10$ scores exceeding $0.997$. As 
$\Theta$ relaxes from $3000$ to $50$, retrieval recall declines for STABLE, StitchedVamana, and FilteredVamana, whereas ACORN's recall improves. Notably, ACORN suffers a severe recall drop at high cardinalities, which may be due to the fact that its subgraph traversal mechanism is optimized for small cardinalities and cannot scale to larger cardinalities. In contrast, STABLE's enhanced heterogeneous semantic awareness enables more precise collaborative measurements. Indeed, STABLE maintains full-precision retrieval (Recall@$10$=$1.0000$) until $\Theta$ is reduced to $1000$ and even when $\Theta$ further relaxes to $50$, its recall decreases by only $0.00298$. This recall drop is one\mbox{-}fifth that of FilteredVamana and one\mbox{-}half that of StitchedVamana, confirming STABLE's superior robustness.

\vspace{0.1cm}
\noindent \textbf{\textcolor{r2}{On Robustness for Query Selectivity.}} 
\textcolor{r2}{In addition to data-side variations (i.e., volume and cardinality), real-world hybrid retrieval scenarios often involve queries with varying degrees of selectivity. To address the concern regarding whether the similarity-based AUTO metric can effectively enforce exact attribute matching under strict filtering conditions, we conduct a stress test on the SIFT-7-3 dataset. Specifically, we adjust the number of active filters ($F$) in queries from 1 to 7, constructing a series of tests ranging from loose filtering ($\approx 33.3\%$ selectivity) to strict filtering ($\approx 0.046\%$ selectivity).}

\textcolor{r2}{As illustrated in Fig.~\ref{fig:stress_test}, we observe the following:
(1) \textit{Guaranteed Precision:} Across all filter settings, STABLE consistently maintains a Recall@10 of nearly 1.0. This empirically confirms that the AUTO metric, despite being a similarity-based measure, functions effectively as a hard constraint mechanism. The heavy penalty for attribute mismatch ensures that nodes failing to meet the ``exact match'' requirement are successfully filtered out, even in high-selectivity scenarios.
(2) \textit{Superior Efficiency:} While baselines like NHQ-K and NHQ-N experience performance degradation as filters increase (due to sparse connectivity in the filtered graph), STABLE maintains high QPS and robustness. Even in the most strictly filtered scenario ($F=7$), STABLE remains in the optimal top-right region of the chart, significantly outperforming partition-based methods (e.g., Milvus) and other graph-based baselines. This demonstrates STABLE's ability to navigate the heterogeneous semantic graph efficiently, regardless of query complexity.}

\subsection{Ablation Analysis (RQ3)}
\label{Ablation Analysis (RQ3)}
To gain further insight into the effectiveness of each key component of STABLE, we conduct detailed ablation experiments on STABLE, and its variants are listed as follows.

\textbf{$\bullet$ w/o~FeatureDis \& w/o~AttributeDis \& w/o~AUTO:} In order to verify the necessity of AUTO, we evaluate the hybrid ANNS performance by removing feature similarity calculation~(\textbf{w/o~FeatureDis}) and attribute consistency evaluation~(\textbf{w/o~AttributeDis}), respectively, and replace AUTO with a simple summation of feature similarity and attribute consistency (\textbf{w/o~AUTO}).

\textbf{$\bullet$ w/o~HSP:} To verify the validity of \textit{Heterogeneous Semantic Prune}, we remove the prune process during the construction of HELP.

\textbf{$\bullet$ w/o~Dynamic\_Routing:} We remove our \textit{Dynamic Heterogeneity Routing} and replace it with the existing state-of-the-art two-stage routing strategy~\cite{nhq} to demonstrate its effectiveness in improving routing efficiency.

\textbf{$\bullet$ w/o~DCR:} To confirm the significance of \textit{Dynamic Coarse Routing}, we remove it and directly perform \textit{Greedy Refinement Routing} in the retrieval process.

As shown in Fig.~\ref{ablation}, we present the retrieval performance results of STABLE's variants on SIFT-7-3, GLOVE-7-3, and CRAWL-7-3, yielding the following insights.
\textbf{(1)} \textbf{w/o~AttributeDis} and \textbf{w/o~FeatureDis} are unable to achieve correct retrieval results, with retrieval accuracy particularly poor (the highest recall of \textbf{w/o~AttributeDis} is only $0.022$ on SIFT-7-3). This indicates that relying solely on feature similarity or attribute consistency to assess heterogeneous semantics neglects part of the semantics, leading to inadequate retrieval results. Although \textbf{w/o~AUTO} uses both feature similarity and attribute consistency for heterogeneous semantic perception, it does not show significant improvement in retrieval performance. This suggests it cannot address similarity magnitude heterogeneity. For example, in the SIFT-7-3 dataset, the feature distance is hundreds of times larger than the attribute distance, causing \textbf{w/o~AUTO} to perform similarly to \textbf{w/o~AttributeDis}. 
\textbf{(2)} Compared to STABLE, the performance of \textbf{w/o~HSP} decreases, indicating that using \textit{Heterogeneous Semantic Prune} to reduce redundant connections in the index structure can assist to reduce computational overhead.
\textbf{(3)} As illustrated in Fig.~\ref{ablation}, although the two-stage routing~(i.e., \textbf{w/o~Dynamic\_Routing}) exhibits satisfactory retrieval performance, our STABLE continues to demonstrate an advantage. This evidence demonstrates the effectiveness of our dynamic routing approach in reducing redundant computations and improving routing efficiency.
\textbf{(4)} The retrieval performance of \textbf{w/o~DCR} decreases, which is rational because the loss of this module makes the routing degenerate into a traditional greedy algorithm. This corroborates the efficacy of the \textit{Dynamic Coarse Routing} in enhancing routing efficiency.

\subsection{Analysis on Index Construction Time (RQ4)}
\label{On Key Hyperparameters Analysis}
\begin{figure}[ht]
	\vspace{-10pt}
 \begin{center}
	\includegraphics[width=\linewidth]{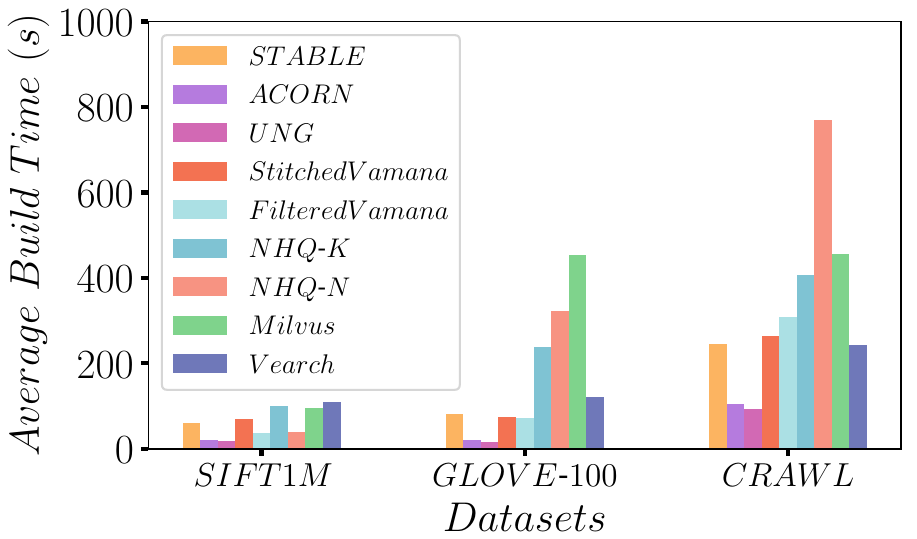}
 \end{center}
  	\vspace{-8pt}
	\caption{Average index build time (s) of hybrid ANNS algorithms on three \textit{1M}-scale dataset categories.}
	\label{IndexTime}
 	\vspace{-8pt}
     \end{figure}

In this section, we conduct an experimental analysis of index construction efficiency, discussing it in terms of construction time. Specifically, we record the average index build time of STABLE and the baselines on the nine datasets that can be classified into three categories in Fig.~\ref{IndexTime}. 

Based on these results, we have the following observations.
\textbf{(1)} For two inverted index-based methods: Vearch and Milvus, Vearch has the median index construction efficiency, while Milvus performs inferior index construction efficiency, shown in Fig.~\ref{IndexTime}. This may be due to the former method performing data storage and index construction simultaneously, and the latter method must build sub-indices within each attribute-guided data partition, resulting in limited building efficiency.
\textbf{(2)} NHQ-K delivers median construction efficiency, 
whereas the construction of NHQ-N reaches inferior efficiency. This indicates that NHQ-K's iterative building strategy is more effective for constructing. However, for NHQ-N, along with the number of nodes~($N$) increasing (node number, SIFT1M:$1,000,000$, GLOVE-100:$1,183,514$, and CRAWL:$1,989,995$), the efficiency of index construction decreases, which is noticeable on CRAWL in Fig.~\ref{IndexTime}. This may be attributed to its incremental construction strategy.
\textbf{(3)} \textcolor{r1}{UNG}, ACORN, FilteredVamana, StitchedVamana, and STABLE show high index\mbox{-}construction efficiency due to the optimized connection strategies. Among them, \textcolor{r1}{UNG} achieves the best construction efficiency, but the experiments above show that the index it constructs still needs to be enhanced for high retrieval accuracy and robustness.
\textbf{(4)} Based on the above analysis, our STABLE achieves robust retrieval performance across datasets while maintaining efficient index construction. It fully demonstrates the superiority of our HELP index.

\subsection{Validation of the Effectiveness of the \texorpdfstring{$\alpha$}{α} Calculation (RQ5)}
\label{appendix:alpha}

In the Section~\ref{sec3}, we propose our \textit{enh\textbf{A}nced heterogeneo\textbf{U}s seman\textbf{T}ic percepti\textbf{O}n (AUTO)} metric to jointly measure feature similarity and attribute consistency, which is calculated by Eq.~\ref{fd}. To obtain the pivotal parameter $\alpha$ in this equation, we first randomly sample $1,000$ nodes from the dataset prior to index construction and compute the average feature distance $\overline{\mathcal{S}_V}$ and the average attribute consistency $\overline{\mathcal{S}_A}$. We then use these two average magnitudes as a reference to set the value of $\alpha$, calculated by Eq.~\ref{alp}.

\begin{figure}
  	\vspace{-10pt}
\begin{center}
    	\includegraphics[width=\linewidth]{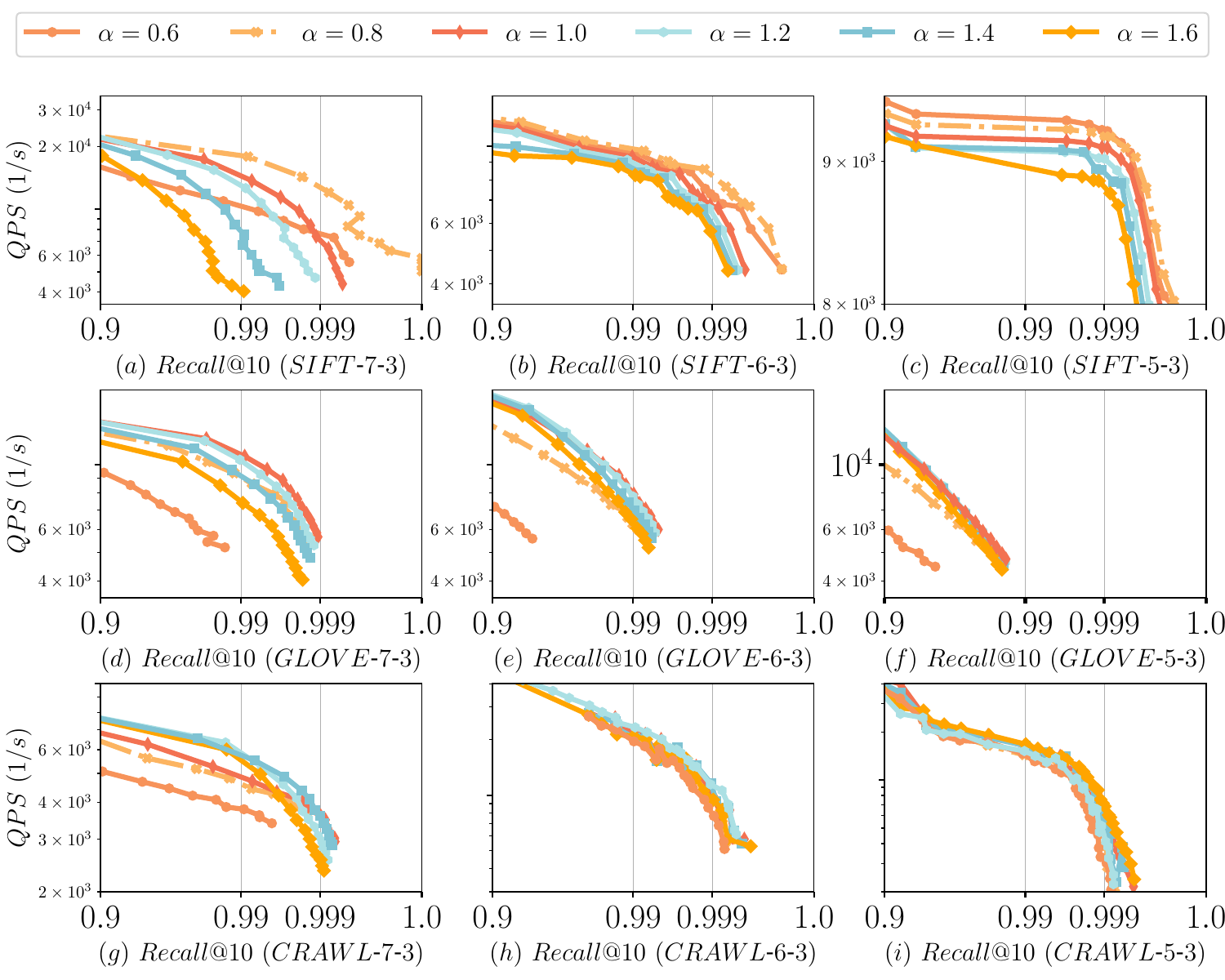}
\end{center}
 \vspace{-6pt}
	\caption{(a)-(i) represent retrieval performance across nine validation datasets for different values of hyper-parameter $\alpha$.}
	\label{alpha}
\end{figure}

\textcolor{r3}{To obtain deeper insight into the robustness of this calculation, we conduct extensive experiments on datasets characterized by diverse distribution patterns. Specifically, these benchmarks encompass the vastly different distributions of SIFT1M, GLOVE-100, and CRAWL (as detailed in TABLE~\ref{datasets}), and cover a wide range of attribute cardinalities ($\Theta$ from 243 to 2187).} \textcolor{r1}{Based on the calculation described above, the derived $\alpha$ values adapt to each dataset are: $\alpha=0.8$ for the SIFT series (5-3, 6-3, 7-3); $\alpha=1.0$ for GLOVE-6-3 and GLOVE-7-3; $\alpha=1.2$ for GLOVE-5-3; $\alpha=1.4$ for CRAWL-6-3 and CRAWL-7-3; and $\alpha=1.6$ for CRAWL-5-3.}

As illustrated in Fig.~\ref{alpha}, the retrieval performance across most of these datasets follows a distinct pattern: increasing initially before declining. This behavior is theoretically grounded: excessively small $\alpha$ values skew the AUTO metric towards attribute consistency, potentially severing links between feature-similar nodes. Conversely, overly large $\alpha$ values diminish attribute awareness, failing to effectively distinguish nodes with identical attributes.

\textcolor{r3}{From Fig.~\ref{alpha}, we also derive two critical observations that substantiate the effectiveness and robustness of our method:}
\textcolor{r3}{1) Adaptability to Diverse Distributions. The empirical optimal $\alpha$ is not static but varies significantly across datasets (ranging from 0.8 on SIFT to 1.6 on CRAWL) due to distinct feature-attribute distributions. Crucially, our calculated $\alpha$ dynamically adjusts to these intrinsic statistics and consistently aligns with the optimal performance peaks across all nine scenarios. This confirms that our method effectively captures underlying data characteristics to achieve optimal results without manual tuning.}
\textcolor{r3}{2) Tolerance to Deviation. Furthermore, the performance landscape surrounding the optimal $\alpha$ manifests as a ``gentle slope'' rather than a precipitous ``cliff''. For instance, on CRAWL-6-3, the QPS remains stable even when $\alpha$ slightly deviates from the calculated optimum of 1.4~(within $[1.2, 1.6]$). This phenomenon indicates that the system is robust to minor statistical fluctuations. Even if the estimated $\alpha$ on an unseen dataset exhibits slight deviations, the performance will not suffer a rapid collapse, providing strong confidence in the practical deployability of STABLE.}


\subsection{\textcolor{r1}{Parameter Sensitivity Analysis (RQ6)}}
\label{appendix:indexsize}

\textcolor{r1}{In this section, we analyze the impact of two key hyperparameters in STABLE: the maximum number of neighbors $\Gamma$ and the pruning threshold $\sigma$. These parameters collectively determine the topology of the HELP index, thereby influencing both index size and retrieval performance.}

\noindent\textbf{\textcolor{r1}{Sensitivity of Pruning Threshold $\sigma$.}}
\textcolor{r1}{The pruning threshold $\sigma$ is critical for filtering redundant edges based on heterogeneous semantic similarity. Unlike $\alpha$, which is data-dependent, $\sigma$ operates on the scale-invariant cosine similarity, theoretically suggesting high stability across datasets.
To verify this and determine the optimal setting, we conduct a systematic sensitivity analysis using a two-stage grid search on SIFT1M, GLOVE-100, and CRAWL datasets.}

\begin{figure*}[t]
  \centering
  \includegraphics[width=\textwidth]{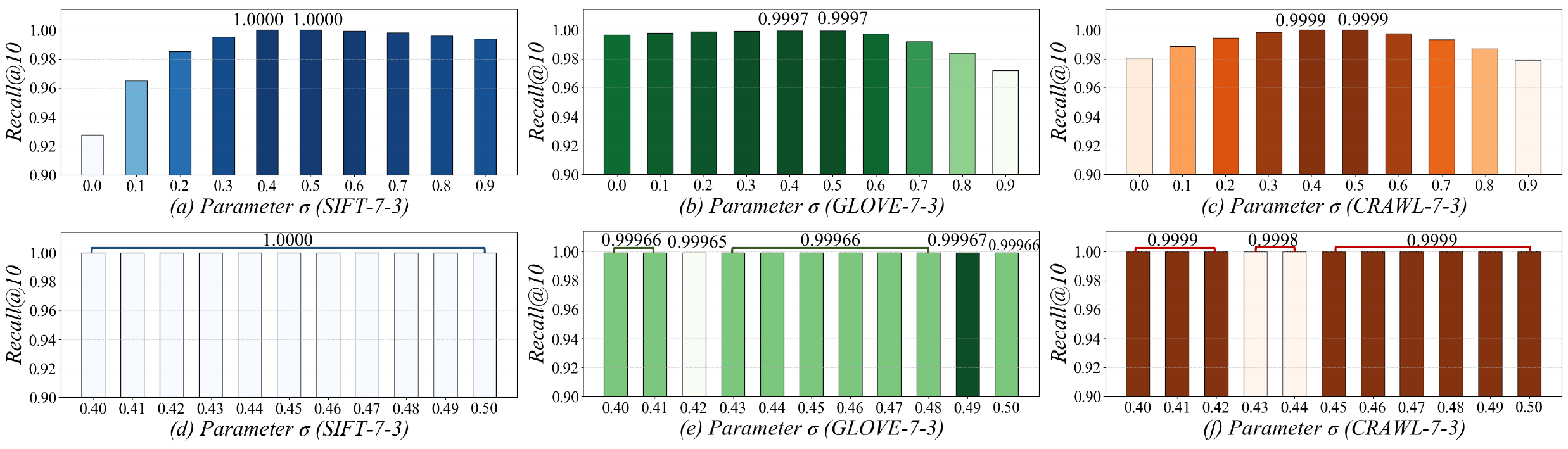} 
  \caption{Sensitivity analysis of the pruning threshold $\sigma$ on three datasets, fixed QPS=2000. (a)-(c) Coarse-grained search ($\sigma \in [0.0, 0.9]$) shows that performance peaks within $[0.4, 0.5]$. (d)-(f) Fine-grained search ($\sigma \in [0.40, 0.50]$) reveals negligible fluctuations ($<0.1\%$) in Recall@10, confirming the parameter's robustness.}
  \label{fig:sigma_sensitivity}
\end{figure*}

\textcolor{r1}{As illustrated in Fig. \ref{fig:sigma_sensitivity} (a)-(c), we first perform a coarse-grained search with a step of 0.1. The results show that Recall@10 reaches a plateau when $\sigma \in [0.4, 0.5]$ across all datasets. 
Specifically, when $\sigma < 0.4$, some valid semantic connections are pruned, reducing search efficiency; when $\sigma > 0.6$, the graph becomes overly dense with redundant edges, causing a drop in recall (e.g., recall drops to $<0.99$ when $\sigma=0.8$ on CRAWL).}

\textcolor{r1}{To pinpoint the optimal value, we conduct a fine-grained search within the $[0.4, 0.5]$ interval with a step of 0.01, as shown in Fig. \ref{fig:sigma_sensitivity} (d)-(f). 
The results exhibit remarkable stability: the fluctuation in Recall@10 is less than $0.1\%$ within this range. 
This empirical evidence validates our hypothesis regarding the stability of $\sigma$. Consequently, STABLE adopts a unified strategy where $\sigma$ is selected within $[0.4, 0.5]$ to ensure robust performance across diverse data distributions.}

\begin{figure}[h]
 \begin{center}
     \includegraphics[width=\linewidth]{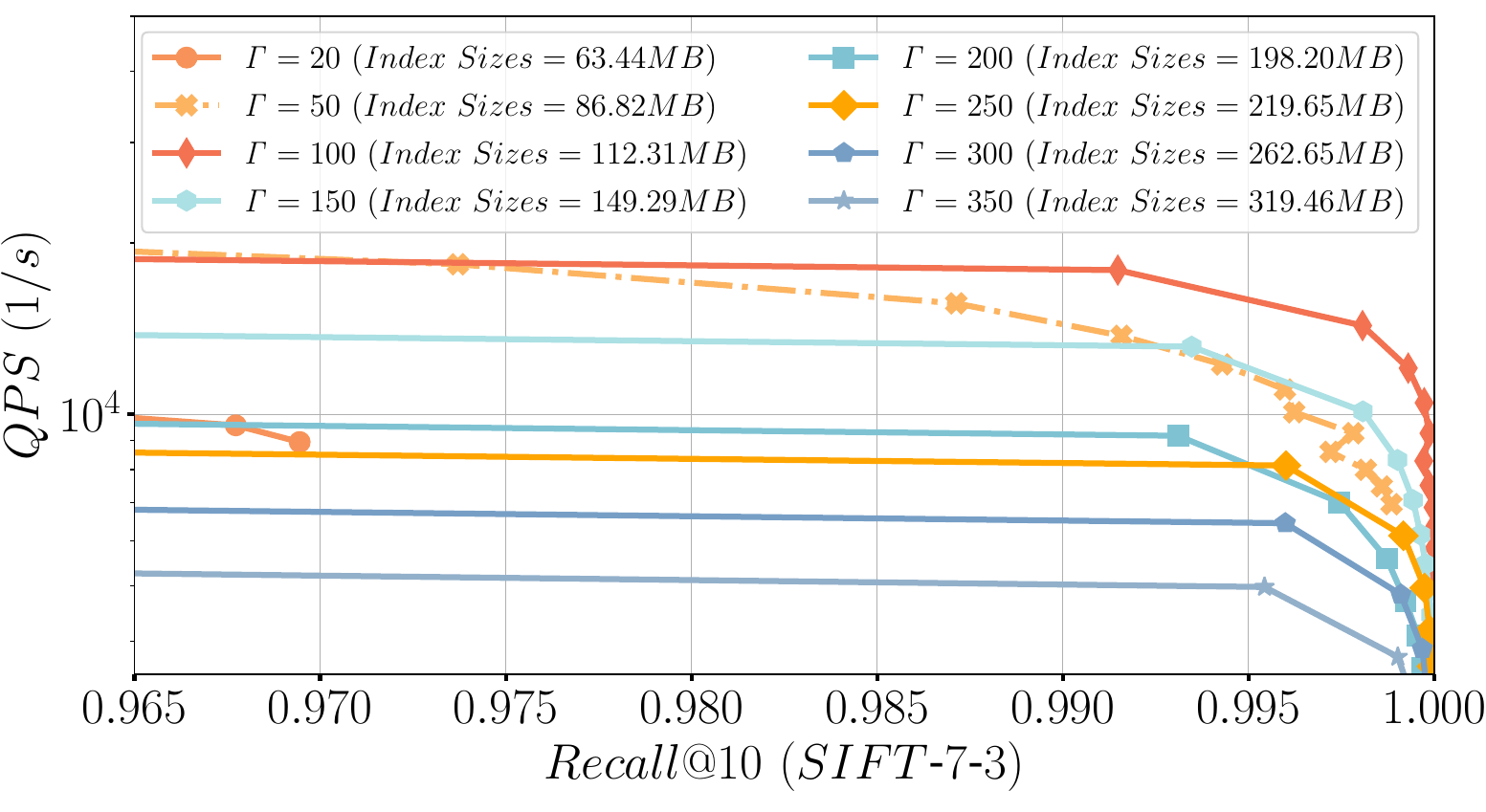}
 \end{center}
  	\vspace{-8pt}
	\caption{Retrieval performance with various $\varGamma$ (i.e., various index sizes) on the SIFT-7-3 dataset.}
	\label{GammaQPS}
\end{figure}

\vspace{1mm}
\noindent\textbf{Impact of Maximum Neighbors $\Gamma$.}
Our graph index size is directly influenced by the key hyperparameter $\varGamma$, which defines the maximum number of neighbors a node can have. To analyze the impact of $\varGamma$ on index size, we conduct experiments comparing ``retrieval performance \textit{vs} $\varGamma$(index size)'' (shown in Fig.~\ref{GammaQPS}) for different values of $\varGamma$ ranging from $20$ to $350$.
From the results, we observe that as $\varGamma$ increases, the index size grows, while retrieval performance initially improves and then declines. This trend can be attributed to two factors: a small $\varGamma$ results in poor connectivity of the graph, while a large $\varGamma$ leads to excessive traversal of certain nodes, both of which degrade retrieval performance. Based on these observations, we set $\varGamma$ to $100$ to achieve optimal performance while maintaining an acceptable index size and construction time.

\subsection{\textcolor{r1}{Impact of SIMD Optimization (RQ7)}}
\textcolor{r1}{To verify that the design of our AUTO metric does not hinder the utilization of SIMD acceleration, we evaluate the retrieval throughput disabling/enabling AVX2 instructions. 
Although AUTO introduces a collaborative measurement of feature similarity and attribute consistency, our design ensures that the computationally intensive feature distance remains fully decoupled, allowing it to be parallelized via explicit AVX2 intrinsics.
}

\textcolor{r1}{We measure the throughput (Queries Per Second, QPS) on the SIFT-7-3 dataset. 
As shown in TABLE~\ref{tab:simd_impact}, the results demonstrate that: \textbf{1)} Enabling AVX2 optimization yields a 2.32$\times$ speedup for STABLE. 
This significant acceleration confirms that our metric design effectively leverages modern CPU vectorization. The performance ceiling is primarily constrained by the memory-bound nature of graph traversal, a phenomenon consistent with Amdahl's Law~\cite{hill2008amdahl}.
\textbf{2)} Compared to the highly optimized Pure Euclidean distance (Pure L2), STABLE incurs a relative overhead of only 4.4\% in terms of QPS. This demonstrates that incorporating additional attribute filtering logic preserves efficient computation without compromising overall high throughput.
}

\textcolor{r1}{Thus, STABLE maintains robust retrieval capabilities with low latency, meeting the stringent efficiency requirements of large-scale industrial deployment.
}

\begin{table}[h]
    \centering
    \caption{Impact of SIMD Optimization on Retrieval Throughput. The throughput is measured in Queries Per Second (QPS), fixed Recall@10=0.999 on SIFT-7-3 dataset.}
    \label{tab:simd_impact}
    \tabcolsep=2pt
    \resizebox{\linewidth}{!}{
    \begin{tabular}{c|c|c|c|c}
        \Xhline{1pt}
        \hline\hline
        \textbf{Method} & \textbf{Instruction Set} & \textbf{Throughput (QPS)} & \textbf{Speedup} & \textbf{Relative Overhead} \\
        \hline
        \multirow{2}{*}{Pure L2 (Baseline)} & Scalar & 4372 & 1.00$\times$ & -- \\
        \cline{2-5} 
         & AVX2 & 10274 & 2.35$\times$ & -- \\
        \hline
        \multirow{2}{*}{\textbf{STABLE (Ours)}} & Scalar & 4234 & 1.00$\times$ & +3.2\% \\
        \cline{2-5}
         & AVX2 & 9822 & 2.32$\times$ & +4.4\% \\
        \hline\hline
            \Xhline{1pt}
    \end{tabular}
    }
\end{table}
\section{CONCLUSIONS}\label{sec5}
In this paper, we analyzed the overlook of the heterogeneity in data distribution and presented a robust heterogeneity-aware hybrid retrieval framework, STABLE. 
To be more specific, we first designed an enhanced heterogeneous semantic perception (AUTO) metric for high-quality collaborative measurement of feature similarity and attribute consistency. Then, based on AUTO, we proposed a heterogeneous semantic relation graph (HELP) index to organize heterogeneous semantic relations. And with the help of the heterogeneous semantic pruning mechanism, HELP can minimize redundant connections to support efficient routing afterward. Furthermore, we introduced a novel dynamic heterogeneity routing to achieve robust and efficient hybrid retrieval. Finally, we performed extensive comparison experiments on three \textit{1M}-scale and two \textit{10M}-scale feature-vector datasets with various attribute cardinalities to demonstrate the effectiveness of our method.

\noindent
\textbf{\textit{Limitations \& Future Work.}} A fundamental limitation of hybrid ANNS methods in \textit{Attribute-Equality} scenario, including our approach, lies in their coarse-grained semantic comprehension of attribute values. These methods can only determine whether two attribute values are strictly identical, but fail to quantify the degree of semantic divergence between them. 
Furthermore, although our framework can support queries involving simple logical operations within a single dimension~(e.g., brand=``A'', style=``casual'', color=``blue'' OR ``yellow'') with a few code adjustments, it exhibits constrained expressiveness when handling more complex compound logic. Consider a composite query requiring: (brand=``A'' OR color=``yellow'') OR (style=``casual'' OR color$\notin$$\{$``green'', ``purple''$\}$). This is also the improvement direction of the existing hybrid ANNS method for more complex applications. 
In the future, we will try to combine knowledge graph embedding and consider introducing complex logic gates to develop a more robust and efficient framework. Moreover, we desire to adopt our method for more applications in information retrieval in the future.


\bibliographystyle{IEEEtran}
\bibliography{references}

@String(AAAI = {AAAI})

@String{Computing = "Computing" }

@String{Computer = "{IEEE} Computer" }

@String{Springer = "Springer-Verlag" }

@STRING{AAAI = "Proc. AAAI"}

@article{du2018spell,
  title={Spell: Online streaming parsing of large unstructured system logs},
  author={Du, Min and Li, Feifei},
  journal={IEEE Transactions on Knowledge and Data Engineering},
  volume={31},
  pages={2213--2227},
  year={2018},
  publisher={IEEE}
}

@article{tekli2016overview,
  title={An overview on xml semantic disambiguation from unstructured text to semi-structured data: Background, applications, and ongoing challenges},
  author={Tekli, Joe},
  journal={IEEE Transactions on Knowledge and Data Engineering},
  volume={28},
  pages={1383--1407},
  year={2016},
  publisher={IEEE}
}

@article{unstructured-datasets,
  title={Mining competitors from large unstructured datasets},
  author={Valkanas, George and Lappas, Theodoros and Gunopulos, Dimitrios},
  journal={IEEE Transactions on Knowledge and Data Engineering},
  volume={29},
  pages={1971--1984},
  year={2017}
}

@article{lu2006hierarchical,
  title={Hierarchical indexing structure for efficient similarity search in video retrieval},
  author={Lu, Hong and Ooi, Beng Chin and Shen, Heng Tao and Xue, Xiangyang},
  journal={IEEE Transactions on Knowledge and data engineering},
  volume={18},
  pages={1544--1559},
  year={2006},
  publisher={IEEE}
}

@article{li2018survey,
  title={A survey of multi-view representation learning},
  author={Li, Yingming and Yang, Ming and Zhang, Zhongfei},
  journal={IEEE Transactions on Knowledge and data engineering},
  volume={31},
  pages={1863--1883},
  year={2018},
  publisher={IEEE}
}

@article{Text-search,
  title={Full text search engine as scalable k-nearest neighbor recommendation system},
  author={Suchal, J{\'a}n and N{\'a}vrat, Pavol},
  journal={in Proceedings of the Artificial Intelligence in Theory and Practice III},
  pages={165--173},
  year={2010},
}

@article{curse-cost,
  title={Approximate nearest neighbor search on high dimensional data—experiments, analyses, and improvement},
  author={Li, Wen and Zhang, Ying and Sun, Yifang and Wang, Wei and Li, Mingjie and Zhang, Wenjie and Lin, Xuemin},
  journal={IEEE Transactions on Knowledge and Data Engineering},
  volume={32},
  pages={1475--1488},
  year={2019}
}

@article{ANN1998,
  title={A quantitative analysis and performance study for similarity-search methods in high-dimensional spaces},
  author={Weber, Roger and Schek, Hans-J{\"o}rg and Blott, Stephen},
  journal={in Proceedings of the International Conference on Very Large Databases},
  pages={194--205},
  year={1998}
}

@article{vearch,
  title={The design and implementation of a real time visual search system on JD E-commerce platform},
  author={Li, Jie and Liu, Haifeng and Gui, Chuanghua and Chen, Jianyu and Ni, Zhenyuan and Wang, Ning and Chen, Yuan},
  journal={in Proceedings of the 19th International Middleware Conference Industry},
  pages={9--16},
  year={2018}
}

@article{ADBV,
  title={Analyticdb-v: A hybrid analytical engine towards query fusion for structured and unstructured data},
  author={Wei, Chuangxian and Wu, Bin and Wang, Sheng and Lou, Renjie and Zhan, Chaoqun and Li, Feifei and Cai, Yuanzhe},
  journal={in Proceedings of the VLDB Endowment},
  pages={3152--3165},
  year={2020}
}

@article{milvus2021,
  title={Milvus: A purpose-built vector data management system},
  author={Wang, Jianguo and Yi, Xiaomeng and Guo, Rentong and Jin, Hai and Xu, Peng and Li, Shengjun and Wang, Xiangyu and Guo, Xiangzhou and Li, Chengming and Xu, Xiaohai and others},
  journal={in Proceedings of the International Conference on Management of Data},
  pages={2614--2627},
  year={2021}
}

@article{nhq,
  title={An Efficient and Robust Framework for Approximate Nearest Neighbor Search with Attribute Constraint},
  author={Wang, Mengzhao and Lv, Lingwei and Xu, Xiaoliang and Wang, Yuxiang and Yue, Qiang and Ni, Jiongkang},
  journal={in Proceedings of International Conference on Neural Information Processing Systems},
  year={2023}
}

@article{filterDiskann,
  title={Filtered-DiskANN: Graph Algorithms for Approximate Nearest Neighbor Search with Filters},
  author={Gollapudi, Siddharth and Karia, Neel and Sivashankar, Varun and Krishnaswamy, Ravishankar and Begwani, Nikit and Raz, Swapnil and Lin, Yiyong and Zhang, Yin and Mahapatro, Neelam and Srinivasan, Premkumar and others},
  journal={in Proceedings of the ACM Web Conference},
  pages={3406--3416},
  year={2023}
}

@article{AttributeCon,
  title={Similarity query processing for high-dimensional data},
  author={Qin, Jianbin and Wang, Wei and Xiao, Chuan and Zhang, Ying},
  journal={in Proceedings of the VLDB Endowment},
  volume={13},
  pages={3437--3440},
  year={2020},
  publisher={Association for Computing Machinery (ACM)}
}

@article{Hash2015,
  title={Query-aware locality-sensitive hashing for approximate nearest neighbor search},
  author={Huang, Qiang and Feng, Jianlin and Zhang, Yikai and Fang, Qiong and Ng, Wilfred},
  journal={in Proceedings of the VLDB Endowment},
  pages={1--12},
  year={2015}
}

@article{ANN-Lazylsh,
  title={Lazylsh: Approximate nearest neighbor search for multiple distance functions with a single index},
  author={Zheng, Yuxin and Guo, Qi and Tung, Anthony KH and Wu, Sai},
  journal={in Proceedings of the International Conference on Management of Data},
  pages={2023--2037},
  year={2016}
}

@article{Hash2020,
  title={iDEC: indexable distance estimating codes for approximate nearest neighbor search},
  author={Gong, Long and Wang, Huayi and Ogihara, Mitsunori and Xu, Jun},
  journal={in Proceedings of the VLDB Endowment},
  volume={13},
  pages={1483--1497},
  year={2020},
  publisher={VLDB Endowment}
}

@article{Hash2020-1,
  title={I/O efficient approximate nearest neighbour search based on learned functions},
  author={Li, Mingjie and Zhang, Ying and Sun, Yifang and Wang, Wei and Tsang, Ivor W and Lin, Xuemin},
  journal={in Proceedings of the IEEE International Conference on Data Engineering},
  pages={289--300},
  year={2020}
}

@article{ANN-Hd,
  title={HD-Index: Pushing the Scalability-Accuracy Boundary for Approximate kNN Search in High-Dimensional Spaces},
  author={Arora, Akhil and Sinha, Sakshi and Kumar, Piyush and Bhattacharya, Arnab},
  journal={Proceedings of the VLDB Endowment},
  volume={11},
  number={8},
  year={2018}
}

@article{b-tree,
  title={iDistance: An adaptive B+-tree based indexing method for nearest neighbor search},
  author={Jagadish, Hosagrahar V and Ooi, Beng Chin and Tan, Kian-Lee and Yu, Cui and Zhang, Rui},
  journal={ACM Transactions on Database Systems},
  volume={30},
  pages={364--397},
  year={2005}
}

@article{Tree2008,
  title={Optimised KD-trees for fast image descriptor matching},
  author={Silpa-Anan, Chanop and Hartley, Richard},
  journal={in Proceedings of the IEEE Conference on Computer Vision and Pattern Recognition},
  pages={1--8},
  year={2008},
}

@article{Tree2014,
  title={Scalable nearest neighbor algorithms for high dimensional data},
  author={Muja, Marius and Lowe, David G},
  journal={IEEE Transactions on pattern analysis and machine intelligence},
  volume={36},
  pages={2227--2240},
  year={2014}
}

@article{Quan2011,
  title={Product quantization for nearest neighbor search},
  author={Jegou, Herve and Douze, Matthijs and Schmid, Cordelia},
  journal={IEEE Transactions on pattern analysis and machine intelligence},
  volume={33},
  pages={117--128},
  year={2010}
}

@article{Quan2020,
  title={Accelerating large-scale inference with anisotropic vector quantization},
  author={Guo, Ruiqi and Sun, Philip and Lindgren, Erik and Geng, Quan and Simcha, David and Chern, Felix and Kumar, Sanjiv},
  journal={in Proceedings of the International Conference on Machine Learning},
  pages={3887--3896},
  year={2020}
}

@article{Quan2015,
  title={Cache locality is not enough: High-performance nearest neighbor search with product quantization fast scan},
  author={Andr{\'e}, Fabien and Kermarrec, Anne-Marie and Le Scouarnec, Nicolas},
  journal={in Proceedings of International Conference on Very Large Data Bases},
  pages={12},
  year={2016}
}

@article{Quan2014,
  title={Optimized product quantization},
  author={Ge, Tiezheng and He, Kaiming and Ke, Qifa and Sun, Jian},
  journal={IEEE Transactions on pattern analysis and machine intelligence},
  volume={36},
  pages={744--755},
  year={2013}
}

@article{NSG,
  title={Fast approximate nearest neighbor search with the navigating spreading-out graph},
  author={Fu, Cong and Xiang, Chao and Wang, Changxu and Cai, Deng},
  journal={in Proceedings of the VLDB Endowment},
  pages={461-474},
  year={2019}
}

@article{SSG,
  title={High dimensional similarity search with satellite system graph: Efficiency, scalability, and unindexed query compatibility},
  author={Fu, Cong and Wang, Changxu and Cai, Deng},
  journal={IEEE Transactions on Pattern Analysis and Machine Intelligence},
  volume={44},
  pages={4139--4150},
  year={2021}
}

@article{DiskAnn,
  title={Diskann: Fast accurate billion-point nearest neighbor search on a single node},
  author={Jayaram Subramanya, Suhas and Devvrit, Fnu and Simhadri, Harsha Vardhan and Krishnawamy, Ravishankar and Kadekodi, Rohan},
  journal={Advances in Neural Information Processing Systems},
  volume={32},
  pages={13766--13776},
  year={2019}
}

@article{FANNG,
  title={Fanng: Fast approximate nearest neighbour graphs},
  author={Harwood, Ben and Drummond, Tom},
  journal={in Proceedings of the IEEE Conference on Computer Vision and Pattern Recognition},
  pages={5713--5722},
  year={2016}
}

@article{HNSW2020,
  title={Efficient and robust approximate nearest neighbor search using hierarchical navigable small world graphs},
  author={Malkov, Yu A and Yashunin, Dmitry A},
  journal={IEEE Transactions on pattern analysis and machine intelligence},
  volume={42},
  pages={824--836},
  year={2018}
}

@article{ann-bench,
  title={ANN-Benchmarks: A benchmarking tool for approximate nearest neighbor algorithms},
  author={Aum{\"u}ller, Martin and Bernhardsson, Erik and Faithfull, Alexander},
  journal={Information Systems},
  volume={87},
  pages={101374},
  year={2020}
}

@article{graphQuality,
  title={Being prepared in a sparse world: the case of KNN graph construction},
  author={Boutet, Antoine and Kermarrec, Anne-Marie and Mittal, Nupur and Ta{\"\i}ani, Fran{\c{c}}ois},
  journal={in Proceedings of the IEEE International Conference on Data Engineering},
  pages={241--252},
  year={2016}
}

@article{nsw,
  title={Approximate nearest neighbor algorithm based on navigable small world graphs},
  author={Malkov, Yury and Ponomarenko, Alexander and Logvinov, Andrey and Krylov, Vladimir},
  journal={Information Systems},
  volume={45},
  pages={61--68},
  year={2014}
}

@article{kgraph,
  title={Efficient k-nearest neighbor graph construction for generic similarity measures},
  author={Dong, Wei and Moses, Charikar and Li, Kai},
  journal={in Proceedings of the International Conference on World wide web},
  pages={577--586},
  year={2011}
}

@article{hu2021coarse,
  title={Coarse-to-fine semantic alignment for cross-modal moment localization},
  author={Hu, Yupeng and Nie, Liqiang and Liu, Meng and Wang, Kun and Wang, Yinglong and Hua, Xian-Sheng},
  journal={IEEE Transactions on Image Processing},
  volume={30},
  pages={5933--5943},
  year={2021}
}

@article{hu2021video,
  title={Video moment localization via deep cross-modal hashing},
  author={Hu, Yupeng and Liu, Meng and Su, Xiaobin and Gao, Zan and Nie, Liqiang},
  journal={IEEE Transactions on Image Processing},
  volume={30},
  pages={4667--4677},
  year={2021}
}

@article{yi2016practical,
  title={Practical approximate k nearest neighbor queries with location and query privacy},
  author={Yi, Xun and Paulet, Russell and Bertino, Elisa and Varadharajan, Vijay},
  journal={IEEE Transactions on Knowledge and Data Engineering},
  volume={28},
  pages={1546--1559},
  year={2016}
}

@article{cai2019revisit,
  title={A revisit of hashing algorithms for approximate nearest neighbor search},
  author={Cai, Deng},
  journal={IEEE Transactions on Knowledge and Data Engineering},
  volume={33},
  pages={2337--2348},
  year={2019}
}

@article{xu2018online,
  title={Online product quantization},
  author={Xu, Donna and Tsang, Ivor W and Zhang, Ying},
  journal={IEEE Transactions on Knowledge and Data Engineering},
  volume={30},
  pages={2185--2198},
  year={2018}
}

@article{ozan2016k,
  title={K-subspaces quantization for approximate nearest neighbor search},
  author={Ozan, Ezgi Can and Kiranyaz, Serkan and Gabbouj, Moncef},
  journal={IEEE Transactions on Knowledge and Data Engineering},
  volume={28},
  pages={1722--1733},
  year={2016}
}

@article{wang2014optimized,
  title={Optimized cartesian k-means},
  author={Wang, Jianfeng and Wang, Jingdong and Song, Jingkuan and Xu, Xin-Shun and Shen, Heng Tao and Li, Shipeng},
  journal={IEEE Transactions on Knowledge and Data Engineering},
  volume={27},
  pages={180--192},
  year={2014}
}

@article{li2002clustering,
  title={Clustering for approximate similarity search in high-dimensional spaces},
  author={Li, Chen and Chang, Edward and Garcia-Molina, Hector and Wiederhold, Gio},
  journal={IEEE Transactions on Knowledge and Data Engineering},
  volume={14},
  pages={792--808},
  year={2002},
  publisher={IEEE}
}

@article{wang2017reverse,
  title={Reverse $ k $ Nearest Neighbor Search over Trajectories},
  author={Wang, Sheng and Bao, Zhifeng and Culpepper, J Shane and Sellis, Timos and Cong, Gao},
  journal={IEEE Transactions on Knowledge and Data Engineering},
  volume={30},
  pages={757--771},
  year={2017},
  publisher={IEEE}
}

@article{wang2021comprehensive,
  title={A comprehensive survey and experimental comparison of graph-based approximate nearest neighbor search},
  author={Wang, Mengzhao and Xu, Xiaoliang and Yue, Qiang and Wang, Yuxiang},
  journal={Proceedings of the VLDB Endowment},
  volume={14},
  number={11},
  pages={1964--1978},
  year={2021},
  publisher={VLDB Endowment}
}

@article{zuo2023arkgraph,
  title={ARKGraph: All-Range Approximate K-Nearest-Neighbor Graph},
  author={Zuo, Chaoji and Deng, Dong},
  journal={Proceedings of the VLDB Endowment},
  volume={16},
  number={10},
  pages={2645--2658},
  year={2023},
  publisher={VLDB Endowment}
}

@article{zhao2023towards,
  title={Towards efficient index construction and approximate nearest neighbor search in high-dimensional spaces},
  author={Zhao, Xi and Tian, Yao and Huang, Kai and Zheng, Bolong and Zhou, Xiaofang},
  journal={Proceedings of the VLDB Endowment},
  volume={16},
  number={8},
  pages={1979--1991},
  year={2023},
  publisher={VLDB Endowment}
}

@inproceedings{DEEP1B,
  title={Efficient indexing of billion-scale datasets of deep descriptors},
  author={Babenko, Artem and Lempitsky, Victor},
  booktitle={Proceedings of the IEEE Conference on Computer Vision and Pattern Recognition},
  pages={2055--2063},
  year={2016}
}

@article{Data-to-text-Generation,
  title={Stylized data-to-text generation: A case study in the e-commerce domain},
  author={Jing, Liqiang and Song, Xuemeng and Lin, Xuming and Zhao, Zhongzhou and Zhou, Wei and Nie, Liqiang},
  journal={ACM Transactions on Information Systems},
  volume={42},
  number={1},
  pages={1--24},
  year={2023},
  publisher={ACM New York, NY, USA}
}

@article{chen2023bias,
  title={Bias and debias in recommender system: A survey and future directions},
  author={Chen, Jiawei and Dong, Hande and Wang, Xiang and Feng, Fuli and Wang, Meng and He, Xiangnan},
  journal={ACM Transactions on Information Systems},
  volume={41},
  number={3},
  pages={1--39},
  year={2023},
  publisher={ACM New York, NY}
}

@article{hu2023semantic,
  title={Semantic collaborative learning for cross-modal moment localization},
  author={Hu, Yupeng and Wang, Kun and Liu, Meng and Tang, Haoyu and Nie, Liqiang},
  journal={ACM Transactions on Information Systems},
  volume={42},
  number={2},
  pages={1--26},
  year={2023},
  publisher={ACM New York, NY, USA}
}

@article{bruch2023approximate,
  title={An approximate algorithm for maximum inner product search over streaming sparse vectors},
  author={Bruch, Sebastian and Nardini, Franco Maria and Ingber, Amir and Liberty, Edo},
  journal={ACM Transactions on Information Systems},
  volume={42},
  number={2},
  pages={1--43},
  year={2023},
  publisher={ACM New York, NY}
}

@article{ahmad2020deep,
  title={A deep learning architecture for psychometric natural language processing},
  author={Ahmad, Faizan and Abbasi, Ahmed and Li, Jingjing and Dobolyi, David G and Netemeyer, Richard G and Clifford, Gari D and Chen, Hsinchun},
  journal={ACM Transactions on Information Systems},
  volume={38},
  number={1},
  pages={1--29},
  year={2020},
  publisher={ACM New York, NY, USA}
}

@article{acorn,
  title={Acorn: Performant and predicate-agnostic search over vector embeddings and structured data},
  author={Patel, Liana and Kraft, Peter and Guestrin, Carlos and Zaharia, Matei},
  journal={Proceedings of the ACM on Management of Data},
  volume={2},
  number={3},
  pages={1--27},
  year={2024},
  publisher={ACM New York, NY, USA}
}

@article{scis-data-2,
  title={From single-to multi-modal remote sensing imagery interpretation: A survey and taxonomy},
  author={Sun, Xian and Tian, Yu and Lu, Wanxuan and Wang, Peijin and Niu, Ruigang and Yu, Hongfeng and Fu, Kun},
  journal={Science China Information Sciences},
  volume={66},
  number={4},
  pages={140301},
  year={2023},
  publisher={Springer}
}

@article{ung,
  title={Navigating labels and vectors: A unified approach to filtered approximate nearest neighbor search},
  author={Cai, Yuzheng and Shi, Jiayang and Chen, Yizhuo and Zheng, Weiguo},
  journal={Proceedings of the ACM on Management of Data},
  volume={2},
  number={6},
  pages={1--27},
  year={2024},
  publisher={ACM New York, NY, USA}
}

@article{HQI,
  title={High-throughput vector similarity search in knowledge graphs},
  author={Mohoney, Jason and Pacaci, Anil and Chowdhury, Shihabur Rahman and Mousavi, Ali and Ilyas, Ihab F and Minhas, Umar Farooq and Pound, Jeffrey and Rekatsinas, Theodoros},
  journal={Proceedings of the ACM on Management of Data},
  volume={1},
  number={2},
  pages={1--25},
  year={2023},
  publisher={ACM New York, NY, USA}
}

@article{lin2025survey,
  title={Survey of Filtered Approximate Nearest Neighbor Search over the Vector-Scalar Hybrid Data},
  author={Lin, Yanjun and Zhang, Kai and He, Zhenying and Jing, Yinan and Wang, X Sean},
  journal={arXiv preprint arXiv:2505.06501},
  year={2025}
}

@article{chronis2025filtered,
  title={Filtered vector search: State-of-the-art and research opportunities},
  author={Chronis, Yannis and Caminal, Helena and Papakonstantinou, Yannis and {\"O}zcan, Fatma and Ailamaki, Anastasia},
  journal={Proceedings of the VLDB Endowment},
  volume={18},
  number={12},
  pages={5488--5492},
  year={2025},
  publisher={VLDB Endowment}
}

@article{ma-nsw,
  title={Multiattribute approximate nearest neighbor search based on navigable small world graph},
  author={Xu, Xiaoliang and Li, Chang and Wang, Yuxiang and Xia, Yixing},
  journal={Concurrency and Computation: Practice and Experience},
  volume={32},
  number={24},
  pages={e5970},
  year={2020},
  publisher={Wiley Online Library}
}

@inproceedings{hqann,
  title={HQANN: Efficient and robust similarity search for hybrid queries with structured and unstructured constraints},
  author={Wu, Wei and He, Junlin and Qiao, Yu and Fu, Guoheng and Liu, Li and Yu, Jin},
  booktitle={Proceedings of the 31st ACM International Conference on Information \& Knowledge Management},
  pages={4580--4584},
  year={2022}
}

@article{hill2008amdahl,
  title={Amdahl's law in the multicore era},
  author={Hill, Mark D and Marty, Michael R},
  journal={Computer},
  volume={41},
  number={7},
  pages={33--38},
  year={2008},
  publisher={IEEE}
}

@inproceedings{retrack,
  title={ReTrack: Evidence-Driven Dual-Stream Directional Anchor Calibration Network for Composed Video Retrieval},
  author={Li, Zixu and Hu, Yupeng and Chen, Zhiwei and Huang, Qinlei and Qiu, Guozhi and Fu, Zhiheng and Liu, Meng},
  booktitle={Proceedings of the AAAI Conference on Artificial Intelligence},
  volume={40},
  number={28},
  pages={23373--23381},
  year={2026}
}

@inproceedings{habit,
  title={HABIT: Chrono-Synergia Robust Progressive Learning Framework for Composed Image Retrieval},
  author={Li, Zixu and Hu, Yupeng and Chen, Zhiwei and Zhang, Shiqi and Huang, Qinlei and Fu, Zhiheng and Wei, Yinwei},
  booktitle={Proceedings of the AAAI Conference on Artificial Intelligence},
  volume={40},
  number={8},
  pages={6762--6770},
  year={2026}
}

@inproceedings{intent,
  title={INTENT: Invariance and Discrimination-aware Noise Mitigation for Robust Composed Image Retrieval},
  author={Chen, Zhiwei and Hu, Yupeng and Fu, Zhiheng and Li, Zixu and Huang, Jiale and Huang, Qinlei and Wei, Yinwei},
  booktitle={Proceedings of the AAAI Conference on Artificial Intelligence},
  volume={40},
  number={25},
  pages={20463--20471},
  year={2026}
}

@article{FineCIR,
  title={FineCIR: Explicit Parsing of Fine-Grained Modification Semantics for Composed Image Retrieval},
  author={Li, Zixu and Fu, Zhiheng and Hu, Yupeng and Chen, Zhiwei and Wen, Haokun and Nie, Liqiang},
  journal={https://arxiv.org/abs/2503.21309},
  year={2025}
}

@inproceedings{OFFSET, 
  title = {OFFSET: Segmentation-based Focus Shift Revision for Composed Image Retrieval}, 
  author = {Chen, Zhiwei and Hu, Yupeng and Li, Zixu and Fu, Zhiheng and Song, Xuemeng and Nie, Liqiang}, 
  booktitle={Proceedings of ACM International Conference on Multimedia},
  pages = {6113–6122}, 
  year = {2025}
}

@inproceedings{HUD, 
  title = {HUD: Hierarchical Uncertainty-Aware Disambiguation Network for Composed Video Retrieval}, 
  author = {Chen, Zhiwei and Hu, Yupeng and Li, Zixu and Fu, Zhiheng and Wen, Haokun and Guan, Weili}, 
  booktitle={Proceedings of ACM International Conference on Multimedia},
  pages = {6143–6152}, 
  year = {2025} 
}

@article{REFINE,
  title={REFINE: Composed Video Retrieval via Shared and Differential Semantics Enhancement},
  author={Hu, Yupeng and Li, Zixu and Chen, Zhiwei and Huang, Qinlei and Fu, Zhiheng and Xu, Mingzhu and Nie, Liqiang},
  journal={ACM Transactions on Multimedia Computing, Communications and Applications},
  year={2026},
  publisher={ACM New York, NY}
}

@inproceedings{encoder,
  title={Encoder: Entity mining and modification relation binding for composed image retrieval},
  author={Li, Zixu and Chen, Zhiwei and Wen, Haokun and Fu, Zhiheng and Hu, Yupeng and Guan, Weili},
  booktitle={Proceedings of the AAAI Conference on Artificial Intelligence},
  volume={39},
  number={5},
  pages={5101--5109},
  year={2025}
}

@inproceedings{median,
  title={Median: Adaptive intermediate-grained aggregation network for composed image retrieval},
  author={Huang, Qinlei and Chen, Zhiwei and Li, Zixu and Wang, Chunxiao and Song, Xuemeng and Hu, Yupeng and Nie, Liqiang},
  booktitle={IEEE International Conference on Acoustics, Speech and Signal Processing},
  pages={1--5},
  year={2025},
  organization={IEEE}
}

@inproceedings{pair,
  title={PAIR: Complementarity-guided Disentanglement for Composed Image Retrieval},
  author={Fu, Zhiheng and Li, Zixu and Chen, Zhiwei and Wang, Chunxiao and Song, Xuemeng and Hu, Yupeng and Nie, Liqiang},
  booktitle={IEEE International Conference on Acoustics, Speech and Signal Processing},
  pages={1--5},
  year={2025},
  organization={IEEE}
}

@misc{MELT,
      title={MELT: Improve Composed Image Retrieval via the Modification Frequentation-Rarity Balance Network}, 
      author={Guozhi Qiu and Zhiwei Chen and Zixu Li and Qinlei Huang and Zhiheng Fu and Xuemeng Song and Yupeng Hu},
      year={2026},
      eprint={2603.29291},
      archivePrefix={arXiv},
      primaryClass={cs.CV}
}

@misc{HINT,
      title={HINT: Composed Image Retrieval with Dual-path Compositional Contextualized Network}, 
      author={Mingyu Zhang and Zixu Li and Zhiwei Chen and Zhiheng Fu and Xiaowei Zhu and Jiajia Nie and Yinwei Wei and Yupeng Hu},
      year={2026},
      eprint={2603.26341},
      archivePrefix={arXiv},
      primaryClass={cs.CV} 
}

\vfill
\end{CJK}
\end{document}